\documentclass[fleqn,usenatbib]{mnras}

\usepackage{newtxtext,newtxmath}

\usepackage[T1]{fontenc}

\DeclareRobustCommand{\VAN}[3]{#2}
\let\VANthebibliography\thebibliography
\def\thebibliography{\DeclareRobustCommand{\VAN}[3]{##3}\VANthebibliography}

\usepackage{graphicx}	
\usepackage{amsmath}	
\usepackage[dvipsnames]{xcolor}
\usepackage{gensymb}
\usepackage{caption}
\usepackage{subcaption}

\newcommand{\rg}{R_{\rm g}}

\graphicspath{{FiguresDyn_CF/}{FiguresR_BB/}}

\title[General relativistic radiation transport]{General relativistic radiation transport: 
Implications for VLBI/EHT observations of AGN discs, winds and jets}

\author[B. Bandyopadhyay et al.]{
Bidisha~Bandyopadhyay,$^{1}$\thanks{E-mail: bidisharia@gmail.com (BB)}
Christian~Fendt,$^{2}$ \thanks{E-mail: fendt@mpia.de (CF)}
Dominik R.G.~Schleicher,$^{1}$
\newauthor
Christos~Vourellis$^{2}$
\\
$^{1}$Departamento de Astronom\'ia, Facultad Ciencias F\'isicas y Matem\'aticas, Universidad de Concepci\'on, \\
Av. Esteban Iturra s/n Barrio Universitario, Casilla 160-C, Concepci\'on, Chile\\
$^{2}$Max Planck Institute for Astronomy, K\"onigstuhl 17, D-69117 Heidelberg, Germany\\
}

\date{Accepted XXX. Received YYY; in original form ZZZ}

\pubyear{2021}

\begin{document}
\label{firstpage}
\pagerange{\pageref{firstpage}--\pageref{lastpage}}
\maketitle

\begin{abstract}
In 2019, the Event Horizon Telescope Collaboration (EHTC) has published the first image of a supermassive black hole (SMBH) obtained via the Very Large Baseline Interferometry (VLBI) technique.
In the future, it is expected that additional and more sensitive VLBI observations will be pursued for other nearby Active Galactic Nuclei (AGN), and it is therefore important to understand which possible features can be expected in such images. 
In this paper, we post-process General Relativistic Magneto-Hydrodynamical (GR-MHD) simulations which include resistivity, thus providing a self-consistent jet formation model, including resistive mass loading of a wind launched from a disc in Keplerian rotation.
The ray-tracing is done using the General Relativistic Ray-Tracing code GRTRANS assuming synchrotron emission. 
We study the appearance of the black hole environment including the accretion disc, winds and jets under a large range of condition, varying black hole mass, accretion rate, spin, inclination angle, disc parameters and observed frequency. 
When we adopt M87-like parameters, we show that we can reproduce a ring-like feature (similar as observed by the EHT) for some of our simulations. 
The latter suggests that such Keplerian disc models thus could be consistent with the observed results. Depending on their masses, accretion rates, spin and the sensitivity of the observation, we note that other SMBHs may show additional features like winds and jets in the observations.

\end{abstract}

\begin{keywords}
black hole physics -- radiative transfer -- methods: numerical -- magnetohydrodynamics -- accretion, accretion discs
\end{keywords}



\section{Introduction}
Accretion towards supermassive black holes is often accompanied by outflows in the form of winds and jets. 
Sources like M87, Sgr A* and many other Active Galactic Nuclei (AGN) are residing inside the centers of nearby 
galaxies. The power of these AGN is driven by accretion flows, showing signatures of outflows and large scale jets 
that are observed across the spectrum.

Extragalactic jets appear as linearly collimated structures being ejected from the central engines with high velocities.
It is commonly accepted that the launching of relativistic jets requires the existence of an accretion disc and magnetic 
fields around a black hole (see e.g. \citet{2015SSRv..191..441H,2019Univ....5...99F}).
The recent observation by the Event Horizon Telescope (EHT) of the innermost jet at 20 micro-arcsecond resolution of the blazar 3C 279  \citep{EHTC2020} has provided an opportunity and hope for observing and understanding the physical processes that lead to the formation of jets in the vicinity of supermassive black holes.

Generally speaking, the jet dynamics is assumed to arise from a combination of magnetic fields and rotational energy. 
The most accepted theoretical frameworks are the Blandford-Znajek (BZ) mechanism \citep{Blandford1977} which states that the relativistic jets can be launched from the magnetosphere of a black hole by extracting its rotational energy,
while the Blandford-Payne (BP) mechanism \citep{Blandford1982} suggests that jets can be formed as a result of magnetocentrifugal 
acceleration of matter from the surface of an accretion flow. 
It is thus important to explore which of these mechanisms corresponds to the observed features of jets from AGN. 
A possible way to investigate the efficiency of each of these processes is to use magnetohydrodynamic (MHD) 
simulations. To launch BZ jets, the MHD equations need to be solved in a general relativistic (GR) framework.
Note, however, that BP jet formation does not consider the evolution of the accretion disc apart from
providing an anchorage in Keplerian rotation. 
In particular, the process of mass loading is not considered.

A physically complete theory, that generalizes the AGN jet-launching mechanism consistently with the observed behavior of the jet, 
is still under development. The general approach is to perform GR-MHD simulations of the close environment of the central black hole and the 
accretion disc to investigate and compare the launching mechanisms of relativistic extragalactic jets. 
A significant number of GR-MHD codes are being used to simulate 
rotating discs around black holes and their resulting outflows \citep{Koide1999, DeVilliers2003, Gammie2003, Noble2006, DelZanna2007, Noble2009,  Bucciantini2013, McKinney2014, Zanotti2015, Porth2017}. 
A recent development is the move to GPU-accelerated GR-MHD simulations \citep{2019arXiv191210192L} that may substantially accelerate
the simulation time.
 
One of the specific features of the simulations in our present study is that we follow the evolution of a thin Keplerian disc 
from the initial setup of simulation. The study of thin discs was pioneered by \citet{Shakura1973} as a purely hydrodynamic approach in the 
non-relativistic limit, while the general relativistic case was formulated by \citet{Novikov1973}. 
Jet launching from Keplerian discs, thus mass loading, using non relativistic resistive MHD was pioneered
by \citet{Casse2002}. 
These simulations apply resistivity in the form of magnetic diffusivity which allows matter to be accreted through 
the magnetic field that threads the disc. 
This then enables the disc material to be loaded on the jet magnetic field, eventually leading the system into an
inflow-outflow structure in a quasi-stationary state. 
This is the approach we also follow in the present paper, applying a prescription similar to \citet{Vourellis2019}
who expanded the physics of the parallel, 3D, conservative, general relativistic MHD (GR-MHD) code HARM3D \citep{Gammie2003,Noble2006,Noble2009} 
by including resistivity in the form of a magnetic diffusivity, following \citet{Bucciantini2013} and
\citet{Qian2017, Qian2018}. 

Radiative transfer calculations are then essential to make a quantitative comparison of accretions flows, jets 
and winds in the vicinity of black holes since it is the radiation from these jets or accretion flows which we observe. 
Thus besides simulating and understanding the physical parameters which drive these flows, it is important to 
understand how radiation transport in a curved spacetime will affect the appearance of such systems in order to
compare them with the observations.
Most of the radiation in such systems is presumably generated near the black hole event horizon, where relativistic 
effects such as Doppler beaming, gravitational redshift  and bending of light become important. 
 
Ray tracing is a convenient method to carry out fully relativistic radiative transfer calculations, where light 
bending is naturally accounted for by taking the rays to be null geodesics in the Kerr metric. The radiative transfer equation can then be solved along geodesics to calculate observed intensities. Early research in this field was initiated to understand the optical appearance of a black hole 
and its environment. 
For instance, \citet{Synge66} investigated the escape of photons from very compact objects and found that these can escape from a slender cone perpendicular to the surface. 
In the limit of a black hole, this cone becomes a perpendicular line. \citet{Luminet1979} studied the appearance of a non-rotating Schwarzschild black hole surrounded by a thin accretion disc.
In this work, effectively they introduced the concept of a black hole shadow, a region towards which no photons are propagating from behind the black hole due to the curvature of space-time; thus casting a ''shadow'' as no light arrives in that region. 
The outer region of the shadow corresponds to the ''photon ring'', which  corresponds to  bound  photon  orbits  around  the  black  hole \citep{Bardeen1973}. 
As a result of lensing, the photons from the lensed inner part of the disc will appear somewhat further outside that region \citep{Luminet1979, Beckwith2005}.
The shape of the shadow in principle will depend on mass, spin and inclination angle of the black hole, as well as potential deviations from General Relativity \citep[e.g.][]{Johannsen2013}. 
For the supermassive black hole in the Galactic Center, its possible appearances including the shadow have been derived by \citet{Falcke2000}.
The method of ray tracing has been used to calculate intensity maps and spectra (see e.g., \citet{Cunningham1975}) and to 
obtain a fit to observed spectra in order to infer parameters such as the spin of the black hole \citep{Davis2006, Li2005, Dauser2010}.  

With the advent of the era of GR-MHD simulations of black hole accretion \citep{DeVilliers2003, Gammie2003}, ray 
tracing has become even more important for the post-processing of these simulations and to analyze their expected appearance to an
observer, including variability properties \citep{Schnittman2006, Noble2009a, Dexter2011} and radiative efficiencies \citep{Noble2011,Kulkarni2011}. 
Many studies have been performed in this regard especially comparing with observations of 
SgrA* \citep{Noble2007, Moscibrodzka2009, Dexter2009, Chan2015, Gold2016} and M87 \citep{Dexter2012,Moscibrodzka2016}. 
Recent investigations in this field have been pursued to define, quantify and understand the origin of the photon ring, lensing rings, black hole shadows and image shadows \citep{Gralla2019,Gralla2020a,Gralla2020b,Gralla2020c,Narayan2019,Bronzwaer2020}. 

In this work our intention is not to quantify the the properties of the black hole shadow or the lensing ring but to explore the
parameter space which determines the visibility of different features (including winds, jets and disc emissions) from similar simulated models.
We apply a more general approach.
Comparing intensity maps and spectra of the simulations will enable us to disentangle the three components of extragalactic jet 
sources -- the accretion disc, the disc wind, and the jet.
In addition to finding the radiation signatures from these distinct components, we explore the physical parameter space which 
strongly influences the visibility of these components. 
The photon trajectories from these components originating from regions close to the black hole will  be significantly
deflected due to the strong field thereby leaving interesting signatures in the intensity maps. Towards the end of the paper, we also examine a more specific case for which we adopt parameters similar to the M87 black hole.

This paper is organized as follows: In section [\ref{sec:DM}] we describe briefly the dynamical modeling and the parameter space that are set up in the rHARM3D 
code to generate winds and jets. Then in section [\ref{sec:RT}] we discuss the importance of the various post processing parameters which affect the total 
synchrotron emission, while in section [\ref{sec:Results}] we discuss the emission spectra and the image features 
that we obtain by varying the different post-processing parameters. In the same section we investigate and compare the various image features that we obtain for the various simulation models for the same set of parameter values. Finally we also discuss about a possible ring-like feature from our simulation models adopting parameter values for M87 as inferred by the EHT observations. We then summarise the inferences of this work in section [\ref{sec:Conclusion}].

\begin{figure*}
 \begin{center}
 \includegraphics[height=1.7in, width=7.3in]{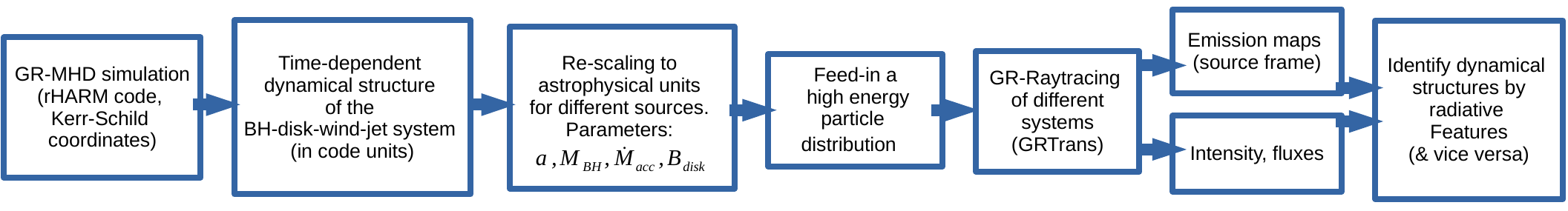}
\caption{Flow diagram of our model approach, combining GR-MHD simulations (applying rHARM) 
and GR radiation transfer (applying GRTRANS) in post-processing.
}
\label{fig:flow-diag}
\end{center}
\end{figure*}

\section{Methodology}
In the present paper,
our main goal is to find radiation signatures for the different components that are hosted by an AGN core. 
These are essentially: 
(i) the close black hole environment, 
(ii) the surrounding accretion disc, 
(iii) the spine jet that is launched by the central rotating black hole via the Blandford-Znajek mechanism, and
(iv) the disc wind that is potentially collimated to a jetted sheath structure surrounding the spine jet, 
     and that is launched from the disc by either the Blandford-Payne magneto-centrifugal process or as a tower 
     jet driven by magnetic pressure.

We aim to explore the parameter space for potential AGN sources to find the conditions under which the 
signatures of accretion, disc winds and jet can be identified in our theoretical approach.
These could be later compared to observed radiation maps, hoping to be able to identify similar features
also in the observed signal.

We proceed as follows (see our flow diagram in Fig.~\ref{fig:flow-diag}).
Firstly, we generate a fiducial set of dynamical models by GR-MHD simulations of thin Keplerian discs around black holes
applying the latest version of our resistive GR-MHD code rHARM3D \citep{Vourellis2019}.
We then scale the models that are derived in GR-MHD code units to a variety of astrophysical sources considering 
different accretion rates or central masses.
Then, the dynamical data obtained by the GR-MHD code are post-processed for ray-tracing using the
publicly available code GRTRANS\footnote{GRTRANS code: https://github.com/jadexter/grtrans} \citep{Dexter2016}, 
which is a fully relativistic code for polarised  radiative transfer via ray tracing in the Kerr metric.
For radiative transfer, we consider synchrotron emission from thermal and non-thermal electrons. The spectra and the images obtained via raytracing for a range of frequencies is then investigated to interpret the various physical processes which leave their imprint on the spectra and images.
We want to stress that our main goal is not to fit certain astrophysical sources, but we aim to find general
radiation patterns of the AGN components. We therefore apply a wide range of parameters for the dynamical modeling that results in a similarly
wide range of spectra or radiation fluxes. Only towards the end of the paper, we also explore a more specific case with M87-like parameters.  As we will see, a variety of dynamical models may lead to similar spectra or overall fluxes,
indicating - potentially - a certain ambiguity when deriving the dynamical parameters for an 
observed source. 

In the following sections we will summarize our approach in more detail.

\subsection{Dynamical modeling} \label{sec:DM}
In order to provide an fiducial set for the magnetohydrodynamic structure of such systems, 
we have performed GR-MHD simulations of the accretion-ejection system.
We have applied the resistive GR-MHD code rHARM3D that has been recently published
\citep{Qian2017, Qian2018, Vourellis2019} 
and that is based on the well-known HARM code 
\citep{Gammie2003,2004ApJ...611..977M, Noble2006,Noble2009}.
We believe that physical resistivity is essential for such simulations as it allows both for a long-term mass loading of 
the {\em disc wind}, and for a smooth accretion process.
This is a consequence of the magnetic diffusivity involved.
Furthermore, it allows to treat reconnection in a way that is physically well defined (in comparison to
reconnection studies applying ideal MHD and relying on numerical resistivity).
The disc wind that is mass loaded from the disc due to physical resistivity, provides an additional
outflow component to the central highly relativistic spine jet, that is launched by the Blandford-Znajek mechanism
\citep{Blandford1977}.
The latter is assumed to be mass loaded by pair production in the disc radiation field; however, this
mechanism is not treated in any GR-MHD simulations, but parametrized by assuming some very low background
density (the so-called floor, see below).
Shearing instabilities between the fast BZ jet and the surround outflow moving with lower speed may also
lead to mass loading by entrainment.

Our GR-MHD simulations are axisymmetric on a spherical grid, meaning that we consider the vector components of all 3 dimensions,
but neglect any derivatives in the $\phi$ direction (sometimes denoted as 2.5~D). 
This limitation is partly due to the exceptionally high CPU request in case of resistive MHD.
However, we believe that our approach is fully sufficient in order to investigate the radiative signatures of the
black hole-disc-wind-jet system.
Of course, we cannot treat any 3D features like instabilities in the disc or in the jets,
nor can we find orbiting substructures in these components.
For our approach we consider such effects of second order only, and thus not as essential for our aim of
disentangling the main AGN components by their radiation features.
{A 3D approach that follows the time evolution of local structural components in the fluid could be applied to auto-correlate
the fluid dynamics with the emerging light curve (see e.g. \citealt{2021PhRvD.103j4038H, 2021CQGra..38l5006C}).
While this would be indeed interesting, it is beyond the scope of our present paper in which we compare the dynamical 
structure and the radiation maps at a certain time step.
Also we expect such an auto-correlation being not compatible with the fast-light approximation.
}

It is worth summarizing the differences of our dynamical model setup to other attempts to model intensity maps and 
spectra from GR-MHD simulations such as e.g. published by \citet{Dexter2012, Moscibrodzka2016, EHTC2019e}.
We work with resistive GR-MHD, that allows for smooth disc accretion and launching of a disc wind.
We start the simulation with a geometrically thin disc in Keplerian rotation, that is thread by a large-scale 
magnetic flux.
Together, this allows to drive strong disc winds (that, on larger spatial scales, are supposed to evolve into 
a jet).
We note that it is in particular
the Keplerian rotation of that disc that may lead to a magneto-centrifugal launching \citep{Blandford1982} of strong disc 
winds.
To include the material of such disc winds in our ray-tracing approach to explore its possible appearance was among our major goals in this paper.

The disc thickness itself plays a minor role for the evolution of the GR-MHD simulation, as the simulations evolve away from the chosen initial state.
While our disc becomes somewhat thicker with time, the initial tori applied in some simulations within the literature typically become thinner. 
This picture would probably change if radiation MHD is considered, treating heating and cooling 
of disc material instantaneously (thus not in post-processing). 
{Such models are becoming available (see e.g. \citealt{2017ApJ...844L..24R,2018ApJ...864..126R,2019MNRAS.486.2873C})
and provide a more consistent evolution of the disc structure.}

In comparison, the above cited simulations start from a thick torus solution, with no net magnetic flux.
While in their application the magneto-rotational instability is the main driver of accretion, in our case
it is the magnetized disc wind that is more efficient in angular momentum removal due to its large lever arm.
A common feature of all GR-MHD applications is the application of a so-called floor density set as a lower density
threshold in order to keep the MHD simulation going.
This numerical necessity however strongly affects the dynamics of the axial Blandford-Znajek jet, 
as the mass flux carried with the jet is directly interrelated with the Lorentz factor (for a given field strength).
Note that the disc wind dynamics is, instead, governed by the mass loading from the disc,
which is self-consistently derived from MHD principles.
Thus, as for all GR-MHD simulations in the literature, the dynamics of the spine jet has to be considered with
some caution.
While the density (and pressure) floor is a necessity in GR-MHD simulations, in our paper we explicitly 
investigate how this choice actually affects the ray-tracing results -- to our knowledge for the first time. 
{However many investigations are being pursued to combat the effect of floor, one of the recent ones as shown by \citet{Kalinani2021arXiv}.}

\begin{figure*}
 \begin{center}
 \includegraphics[height=2.5in,width=7.3in]{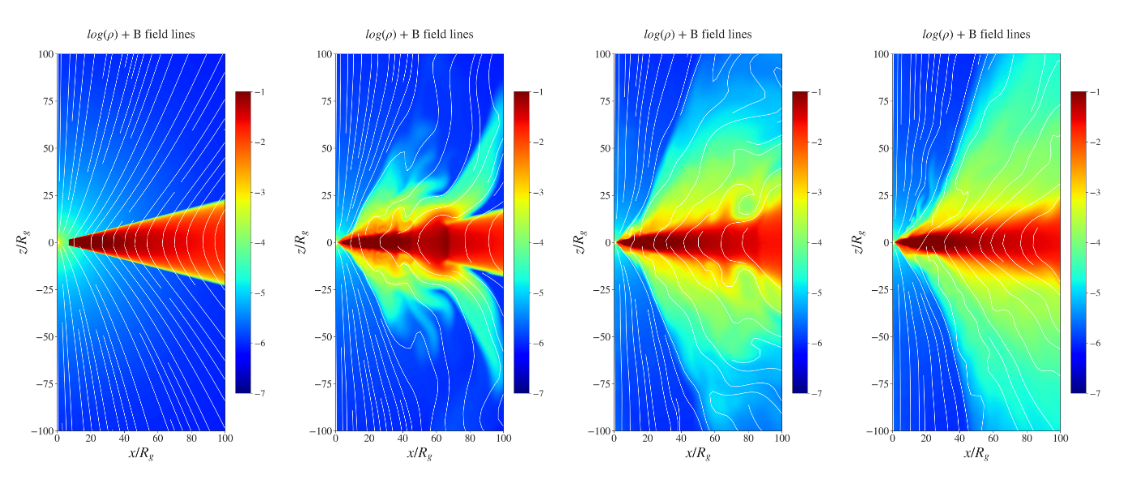}
\caption{Reference run {\em 20EF} of our GR-MHD simulations.
Shown is the evolution of the density (colors, in log10 scale) and magnetic field (poloidal field lines, white) for 
four different time steps, $t=0$, $t=2000$, $t=5000$, and $t=10000$ (from left to right).
The panels show the highly resolved area of $r<100$.
Lengths are normalized in gravitational radii $R_{\rm g}$, while the time unit is $t_{\rm g}$. }
\label{fig:sim20_t-evol}
\end{center}
\end{figure*}

Our simulations follow the overall setup that we have recently published \citep{Vourellis2019} and for
a detailed description we refer to that publication. 
The code solves the exactly same equations as described in \citet{Vourellis2019}.
The output of our GR-MHD simulations comes in normalized units where  $G=c=1$
for the gravitational constant $G$ and the speed of light $c$.
In particular, all 
length scales shown here are in gravitational radii $R_{\rm g} = GM/c^2$, 
velocities are normalized to the speed of light,
while time is measured in units of the light travel time $t_{\rm g} = GM/c^3$.
The dimensionless spin parameter is $a = Jc/GM^2$ for a black hole with mass $M$ and angular momentum $J$.
All physical variables need to be properly re-scaled for post-processing of radiation transfer.

In the following, we briefly summarize our approach for the dynamical modeling of GR-MHD jets in this project.

\subsubsection{Initial and boundary conditions}
We choose a grid stretching that allows for a sufficiently high resolution close to the black hole, 
and on the other hand provides the option to move the outer domain boundary far from the central 
area, to avoid any feedback from the outer boundary to the central area.
Our numerical grid size is 512x255 grid cells applying spherical coordinates. 
The outer boundary is located at $r_{\textrm {out}} = 1000 \rg$. 
The grid is stretched in the radial direction, providing a higher resolution close to the black hole, and a lower, 
but still sufficient resolution at larger radii.
In order to investigate the inflow-outflow structure of the disc wind - the actual {\em launching} of the outflow - we also 
apply a (sinusoidal) grid refinement in the disc area, resolving the (initial) disc scale height with about 20 grid cells.
The choice of resolution is also constrained by the application of physical resistivity, as the numerical time step
for diffusivity decreases with $\propto 1/(\Delta r)^2$.

The initial condition is that of a Keplerian thin disc considering a vertical Gaussian density and pressure
distribution with an initial scale height $h = 0.1 r$.
We apply an ideal gas law with an initial polytropic index $\gamma_{\rm gas} = 5/3$, 
and an entropy parameter for the disc gas $\kappa_{\rm disc} \equiv \rho/ P^\gamma_{\rm gas} = 10^{-3}$.
The disc carries a large-scale vertical magnetic flux that is normalized with respect to the disc pressure maximum
by the parameter plasma-$\beta$.

The disc-black hole system is initially embedded in a {"}corona{"} of low density, 
a factor $10^{-3}$ less dense than the accretion disc.
However, this initial corona is not in equilibrium.
It will be partly washed out of the computational domain by the disc wind and the central spine jet.
It may collapse towards the black hole where no wind or jet is present.
While the corona is of low density, its pressure (or internal energy), is relatively
large. 
The gas pressure of this hot gas is large enough to balance the magnetic pressure of the large-scale field.

The boundary conditions are following the prescriptions of HARM, thus considering pure outflows in the radial direction 
(radially out of the grid, or into the black hole) and axisymmetric conditions along the rotational axis.

As MHD codes can hardly treat very low densities, a density (and pressure) floor has to be provided. 
Thus, when the simulated density (or pressure) falls below the local floor value it is replaced by the floor value. 
We note here that while the disc wind dynamics is governed by the mass loading from the disc,
the spine jet dynamics (Lorentz factor, mass flux) depends heavily on the applied floor model.
The latter is a common feature of all GR-MHD simulations published so far.

The physical resistivity $\eta$ is fixed in space and time.
The diffusivity is expected to be turbulent in nature, thus resulting from an $\alpha$-turbulence, 
as it is orders of magnitude larger than a micro-physical resistivity \citep{Shakura1973}.
We apply a disc diffusivity that follows an exponential profile in the vertical direction and a certain
radial profile that is connected to the disc sound speed.
This is motivated by theoretical estimates and direct numerical simulations connecting 
the turbulent disc diffusivity to the sound speed of the disc gas (see \citet{Vourellis2019} and references therein). 
We apply a maximum disc diffusivity which is a factor of ten times higher than that of the background 
diffusivity of $\eta_{\rm back} = 10^{-3}$ (in code units).

We also note that the existence of resistivity suppresses the evolution of the disc magneto-rotational 
instability \citep{1991ApJ...376..214B}.
This has been tested in detail GR-MHD in \citet{Qian2017}.
Without the magneto-rotational instability being present in the disc, the removal of disc angular momentum is done by 
the stresses of the large scale magnetic field.

While the initial conditions and the resistivity distribution are well motivated astrophysically, 
here our parameter choice is also driven by our aim to consider simulations that exhibit  the 
different system components (i)-(iv) (see Sect.~1) that may exist in AGN cores.

\begin{figure}
 \begin{center}
\includegraphics[height=2.5in,width=3.5in]{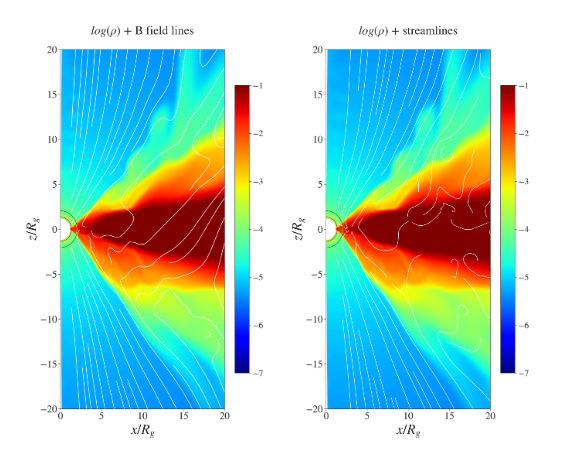}\\
\includegraphics[height=2.5in,width=3.5in]{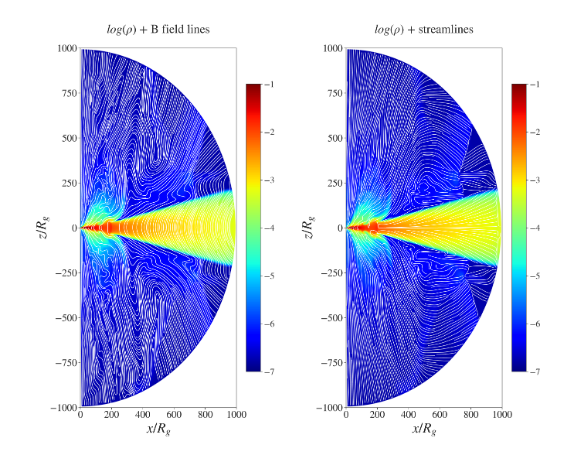}
\caption{Reference run {\em 20EF} of our GR-MHD simulations.
The panels show the highly resolved inner area of $r<20$ top),
and the overall simulation domain of $r<1000$ (bottom).
Shown is the density distribution (colors, in log10 scale) overlaid with the poloidal magnetic (white field lines, top),
and the density distribution (in log10 scale, overlaid with poloidal stream lines (white, bottom) at time $t=10000$.
Lengths are normalized in gravitational radii $R_{\rm g}$, while the time unit is $t_{\rm g}$.
The colored circles (top panels) indicate the horizon (red), the ergosphere (yellow) and the radius of the marginally stable 
orbit (dark green) located in the equatorial plane.
}
\label{fig:sim20_sml-lrg}
\end{center}
\end{figure}

\subsubsection{A reference simulation}
We first briefly discuss the evolution and the resulting dynamics of our reference simulation.
This is simulation {\em 20EF} which applies similar simulation parameters as \citet{Vourellis2019}. 
Different from the original simulation, here we apply a lower background diffusivity, ten times lower than 
the disc diffusivity, in addition to the diffusivity in the disc.
Effectively, the level of background diffusivity leads to a smoother evolution of the very inner parts of the 
simulation domain located within the innermost stable circular orbit (ISCO) and close to the horizon.
This allows us to find a more steady accretion-outflow structure.
We like to note that such a structure is also more consistent with the approach of post-processing ray-tracing. 
Strongly time-dependent phenomena in this inner area which involve high velocities may violate a 
ray-tracing procedure that is - as usual - done under the so-called {"}fast-light approximation{"}.
The fast-light approximation assumes that all phenomena move or change with a speed slower than the speed 
of light. 
The latter defines the time scale for radiation transfer.
When rapidly moving or varying features should be considered, time retardation needs to be taken into account.

In Figure~\ref{fig:sim20_t-evol} we show the time evolution of our reference simulation.
From an initially Keplerian disc that is embedded in a static corona, a disc wind emerges from an increasing 
area of the disc surface.
In Figure~[\ref{fig:sim20_sml-lrg}] we show the dynamical structure of the flow at small and large scales.
Ray-tracing can be done over the whole grid of the simulation domain, thus considering the emission and
absorption of all material moving around till $r = 1000 R_{\rm g}$.

\begin{figure}
 \begin{center}
\includegraphics[height=2.5in,width=3.5in]{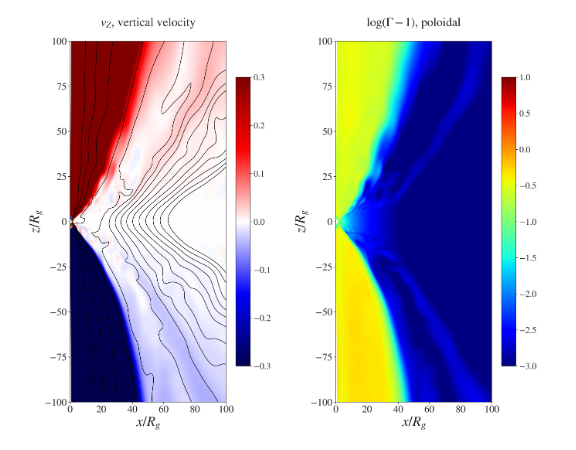}
\caption{Reference run {\em 20EF} of our GR-MHD simulations.
Shown is the distribution of exemplary  MHD variables at time $t=10000$, i.e.
the vertical speed (in linear scale), 
and the Lorentz factor $\Gamma$, shown as $(\Gamma-1)$ (in log10 scale).
The panels show the highly resolved area of $r<100$.
Lengths are normalized in gravitational radii $R_{\rm g}$, while the time unit is $t_{\rm g}$. 
}
\label{fig:sim20_mhd}
\end{center}
\end{figure}

Simulation run {\em 20EF} also produces a high-velocity Blandford-Znajek spine jet (best visible in Figure~[\ref{fig:sim20_mhd}]).
This is mainly due to the low floor density applied to this simulation.
Note that the density of the disc wind is substantially higher than that of the central spine jet
(see the green and dark blue colors highlighting high and low density areas).

\subsubsection{Exploring the parameter space}
In order to investigate the radiation features of systems with different strengths of the individual components,
we have run simulations where we have changed the major simulation parameters compared to {\em 20EF}. 
An overview over our parameter runs is given in Table~\ref{tab:para_dynamics}.

Simulation {\em 21EF} applies a (ten times) higher floor density, thus a higher mass loading of the Blandford-Znajek jet, 
which in turn leads to a lower velocity of the jet. 
Simulation run {\em 22EF} applies a Kerr parameter $a=0$ for which a Blandford-Znajek jet cannot be launched at all.
Run {\em 23EF} applies an even higher floor density (another factor ten), resulting in a super-Alfv\'enic area close to
the black hole that also
prohibits the launching of a Blandford-Znajek jet (see \citealt{2009MNRAS.397.1153K}).
Simulation {\em 24EF} applies a lower plasma-$\beta$ (factor ten) for the initial setup, resulting in a 
stronger magnetic flux, and therefore in
a more effective launching of both the Blandford-Znajek jet and the disc wind.

Run {\em 26EF} applies an even lower plasma-$\beta$ (another factor ten) for the initial setup than {\em 24EF}.
This implies a much faster MHD evolution of the system that will reach a similar dynamical state at comparatively
earlier times. 
We therefore stopped the simulation at $t=4000$ compared to the $t=10000$ simulation time of the other
simulations.

\begin{table*} 
\begin{center}
\caption{Characteristic parameters of our simulation runs. Shown are 
the simulation run ID, 
the maximum disc magnetic diffusivity $\eta_{\rm max}$, 
the initial disc plasma beta $\beta_0$, 
the density ratio between initial disc and coronal density ${\rho_{\rm cor}}/{\rho_{\rm disc}}$, 
the maximum floor density $\rho_{\rm flr}$ and 
internal energy $u_{\rm flr}$,  
the Kerr parameter of the black hole $a$,
the radius of the ISCO $R_{\rm isco}$,
and specific comments on the simulation runs, respectively.
A background diffusivity of $\eta_{\rm back}=10^{-3}$ is applied.}
\label{tab:para_dynamics}
\begin{tabular}{ccccccccl}
\hline
\hline
\noalign{\smallskip}

ID   & $\eta_{\rm max}$ & $\beta_0$ & $\frac{\rho_{\rm cor}}{\rho_{\rm disc}}$ & 
                  $\rho_{\rm flr}$ & $u_{\rm flr}$  & $a$ & $R_{\rm isco}$ & comments\\
\noalign{\smallskip}
\hline
\noalign{\smallskip}
\noalign{\smallskip}
 20EF  & $10^{-2}$ & 10  & $10^{-3}$ & $10^{-5}$ & $10^{-8}$  & 0.9375 & $2~R_{\rm g}$& reference run, similar to \citet{Vourellis2019} \\
 21EF  & $10^{-2}$ & 10  & $10^{-3}$ & $10^{-4}$ & $10^{-6}$  & 0.9375 & $2~R_{\rm g}$ & as 20EF, floor density higher\\
 22EF  & $10^{-2}$ & 10  & $10^{-3}$ & $10^{-4}$ & $10^{-6}$  & 0.0    & $6~R_{\rm g}$ & as 20EF, $a=0$, infall to BH, disc wind \\
 23EF  & $10^{-2}$ & 10  & $10^{-3}$ & $10^{-3}$ & $10^{-5}$  & 0.9375 & $2~R_{\rm g}$ & as 21EF, floor even higher, no BZ anymore \\
 24EF  & $10^{-2}$ &  1  & $10^{-3}$ & $10^{-3}$ & $10^{-5}$  & 0.9375 & $2~R_{\rm g}$ & as 21EF, $\beta_0$ lower, stronger magn. field \\
 26EF  & $10^{-2}$ & 0.1 & $10^{-3}$ & $10^{-3}$ & $10^{-5}$  & 0.9375 & $2~R_{\rm g}$ & as 24EF, $\beta_0$ even lower, even stronger magn. field \\
  \noalign{\smallskip}
 \hline
 \noalign{\smallskip}
 \end{tabular}
 \end{center}
\end{table*}

Besides the obvious impact of choosing a lower plasma-$\beta$, namely the higher magnetic field strength that
naturally would increase the synchrotron power, we show the influence of the parameter choice on the outflow
velocity and density.

In comparison to run {\em 20EF}, by increasing the density and pressure floor values ({\em 21EF}, see Figure~\ref{fig:para_vel}) 
we find a lower jet speed, but also an increased jet luminosity (see below).
Considering energy transformation from magnetic energy to kinetic energy for jet acceleration (for a fixed
black hole rotation rate), this makes sense. 
The low-density material can be accelerated to higher speed.

Further increasing the floor values (run {\em 23EF}) would in principle further increase 
the jet power, however,
as the jet launching region (the black hole ergosphere) becomes super-Alfv\'enic for such high densities,
no BZ jet is launched (see Figure~\ref{fig:para_vel}), instead we obtain a large-scale infall towards the black hole.
This nicely agrees with the prediction of \citealt{2009MNRAS.397.1153K} that BZ jets can only be launched sub--Alfv\'enically.

Obviously, simulation {\em 22EF} that corresponds to a non-rotating Schwarzschild black hole does not produce BZ jets.
However, an interesting prospect may be derived from the velocity maps of this simulation.
Instead of a high velocity BZ jet, we observe high velocity mass infall towards the black hole 
(see Figure~\ref{fig:para_vel}).
The infall velocity reaches intermediate Lorentz factors, similar to the BZ jet speed in the other parameter runs.
Therefore, assuming a similar radiation efficiency of the infalling material as for the
BZ jet in the other simulations, when looking onto the system we may hypothesize whether we may detect
highly Doppler boosted radiation that originates from the {\em far side of the black hole.}
In other words, the idea here was that material that is falling into the BH from the far side of the BH is Doppler-boosted 
towards the observer, just as the jet material that is ejected from the BH towards the observer. 
So in this respect, there might be a potential difficulty to distinguish between inflow and outflow close to the black hole.
On the other hand, the infall towards the BH at the near side of the BH moves away from the observer, is thus 
de-boosted and thus unobservable.

\begin{figure*}
 \begin{center}
\includegraphics[height=2.5in,width=7.0in]{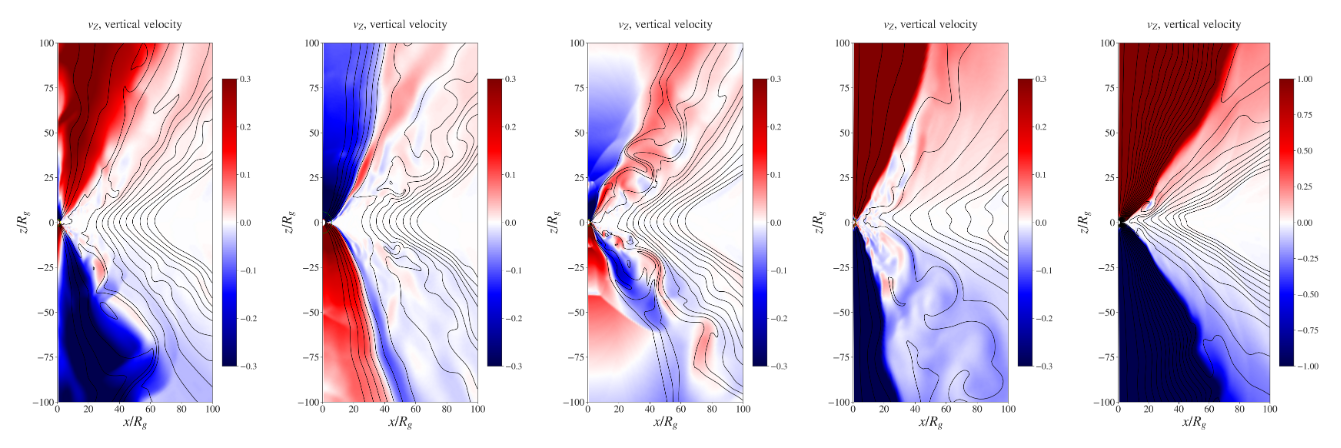}
\caption{Comparison of the flow velocities for the parameter runs.
From left to right we show the vertical velocity (colors, in log10 scale) for
simulation runs 
{\em 21EF}, {\em 22EF}, {\em 23EF}, {\em 24EF}, {\em 26EF},
all at time $t=10000$, except {\em 26EF} shown at $t=4000$.
Overlaid on the velocity color-coding are contours of the vector potential (tracing the
poloidal magnetic field lines), with arbitrary flux levels.
Lengths are normalized in gravitational radii $R_{\rm g}$, while the time unit is $t_{\rm g}$. 
}
\label{fig:para_vel}
\end{center}
\end{figure*}

Since the highest infall speed is reached close to the black hole, the radiation it emits would also be heavily
affected by gravitational lensing -- in contrast to the signal from a BZ jet.

In comparison, simulations with rather high floor values and low plasma-$\beta$ (runs {\em 24EF} and {\em 26EF}) 
are expected to produce the strongest signal for the BZ jet.
Naturally, we expect a higher density ({"}more mass{"}) producing more radiation.
In addition, synchrotron radiation strongly depends on the magnetic field strength.
The jet speed is essential for beaming the radiation.
A strong BZ jet is indeed visible (see comparison in Fig.~\ref{fig:para_vel}).
We note that run {\em 26EF} with the strongest magnetic field has a larger opening angle, probably due to
the large magnetic pressure involved.
The strong field also leads to a larger jet speed compared to run {\em 24EF}.

The outflow speed essentially determines the boosting of radiation seen under small inclination angles in 
addition to the density, temperature and the magnetic field which primarily determine the radiation field.
The electron density, temperature and the magnetic field strength are directly considered in the thermal 
synchrotron formula (the non-thermal distribution function 
{used in this work does not explicitly depend on the temperature}).
A comparison of the mass density for the different parameter runs is shown in Figures~\ref{fig:para_dens}.
For completeness we also show comparative plots for the internal energy (defining the 
gas temperature, Figure~\ref{fig:para_ener}), and the magnetic energy (Figure~\ref{fig:para_btot}).
 
We note that the disc structure remains more or less the same for all simulation runs.
The disc scale increases somewhat from its initial value $h = 0.1~r$ to say $h/r = \simeq 0.2$,
but the disc remains thin.
A stronger magnetic field may provide a higher rate of angular momentum removal from the disc
due to the larger lever arm.
As a result the accretion rate may increase.
At the same time, accretion can be lowered due to the outward curvature force of the bent magnetic field.
Overall, for our approach, these effects are expected to cancel, and we may safely assume the
same accretion rates (in code units) for all simulation runs.

The accretion rate derived from the GR-MHD simulations in code units is $\overline{\dot{M}} \simeq 0.1$,
which can be re-scaled to astrophysical units assuming typical AGN accretion rates of 
\begin{equation}
    \dot{M} \equiv \rho_0  R_{\text g}^2  {\text c} \overline{\dot{M}} \simeq 10^{-3}\,M_{\odot}\text{yr}^{-1}.
\end{equation}
With this we may constrain the (maximum) disc density to
\begin{equation}
\rho_0 = 4\times 10^{-15}\frac{\text{g}}{\text{cm}^3}  \left[ \frac{ \dot{M} }{10^{-3}\,M_{\odot}/\text{yr}} \right] 
                                    \left[ \frac{ M_{\text{BH}} }{ 5\times 10^9 M_{\odot} } \right]^{-2}
                                    \left[ \frac{ \overline{\dot{M} } }{ 0.1 } \right]^{-1}\!\!\!\!
\end{equation}        
\citep{Vourellis2019}.
The re-scaled astrophysical density is $\rho = \rho_0 \bar{\rho}$ where $\bar{\rho}$ in code units follows from 
our simulations.
For illustration, for the example of M87 we may use the values provided by the EHT collaboration \citep{EHTC2019e}
proposing an accretion rate of 
$\dot{M}\sim 2.7\times 10^{-3}\,M_{\sun}\rm yr^{-1}$ 
assuming a black hole mass of $M = 6.2\times10^9\,M_{\sun}$.
In this paper, however, in order to perform the ray tracing on physically scaled variables, we follow a different
approach. 
We will re-normalize the GR-MHD code units by different central masses and accretion rates (see below),
which will allow us in principle to connect or numerical models to a variety of AGN sources.

\subsection{Post-processing with radiative transfer} \label{sec:RT}
Together with the dynamical modelling, the radiation transport mechanism of the flux produced in these
systems plays a key role in determining the spectral shape and appearance of the post-processed systems which is essential to compare the ray-traced simulation results with observations.
Due to strong GR effects in the vicinity of black holes, the photons emitted from those regions follow 
trajectories in the space-time curved by the black hole, which give rise to important observational signatures 
like the lensed photon ring observed in case of M87 \citep{EHTC2019a}. 

GR-MHD codes like HARM \citep{Gammie2003} apply normalized units where $G=c=M=1$, and thus transforming the setup to physical
units can result in a completely different physical picture depending on the various physical parameters used during the post-processing. 
To understand the behaviour of the simulated GR-MHD data in different physical regimes, it is important to include the 
radiative transport equations in a GR frame work to post-processes the data from the simulations in proper physical 
units. 
The goal of a ray-tracing code is to calculate the observed intensity on locations (pixels) of an 
observer’s camera for a given model of emission and absorption. 
In order to achieve this, here we use the publicly available general relativistic radiative transport code GRTRANS \citep{Dexter2016}. 

In GRTRANS, the Boyer-Lindquist coordinates of the photon trajectories are calculated from the observer towards the black 
hole (i.e. the rays are traced) for each pixel in the camera. 
The observed polarisation basis is parallel-transported into the fluid frame. 
Then the local emission and absorption properties at each location are calculated.
Finally the radiative transfer equations for the given emission and absorption are solved along those rays. 
For details on the working of the code, we refer to the original paper by \citet{Dexter2016}.

The code also considers various types of emission that can be included in the treatment - thermal and non-thermal synchrotron,
bremstrahlung, and blackbody radiation. 
The primary aim of this work is to observe the emission features from disc winds and jets very close to the 
central black hole with the high resolution imaging facilities available with Very Large Baseline Interferometry 
(VLBI) techniques such as the EHT, which is possible currently only through radio observations. These jet and wind dominated regions are highly magnetised and thus one can assume synchrotron emission as the dominant emission mechanism in these regions \citep{Yuan2014}. We thus consider only synchrotron (both thermal and non-thermal) emission . 
Understanding the complete spectra including other radiation mechanisms such as bremstrahlung emission and Compton scattering
(such as done for many 1D steady state advection dominated accretion flow models (ADAF) \citep{Bandyopadhyay2019, Nemmen2014, Manmoto1997}.)
is beyond the scope of this current work. Other literature (for reference see \citealt{Moscibrodzka2016,EHTC2019a} etc.) where the spectra, for sources of interest to the EHT, are obtained usually considering the synchrotron emission and its Compton scattering using the GRMONTY code \citep{Dolence2009} but in this work the consideration of synchrotron emission is sufficient as we are only interested in the approximate appearance of different components of the AGN.  
All the spectra or intensity maps thus shown here are generated via pure synchrotron emission by thermal , non-thermal or both the types of electron distribution functions.

For the ray-tracing procedures we use the dynamical model {\em 21EF} as an example to test and compare the effect of various physical parameters. We later fix our choice of all parameter values to compare the emission from our different simulation models. In the following we discuss the important input parameters which are involved in post processing of the simulated HARM 
data that affect the radiative transport mechanism and hence the synchrotron spectrum.

\begin{figure*}
 \begin{center}
 \includegraphics[height=2.2in,width=3.1in]{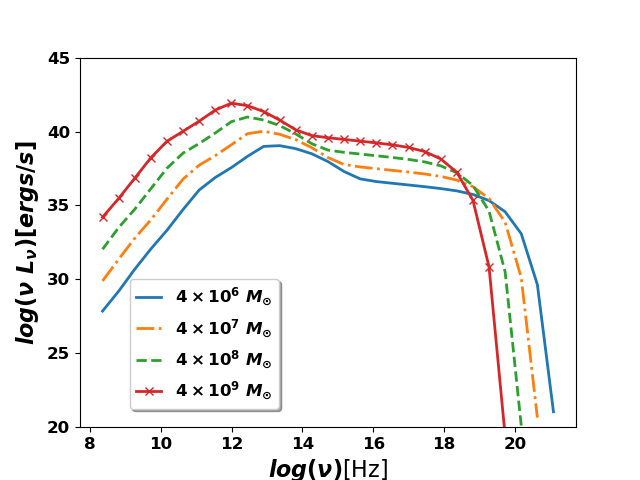}
   \includegraphics[height=2.2in,width=3.1in]{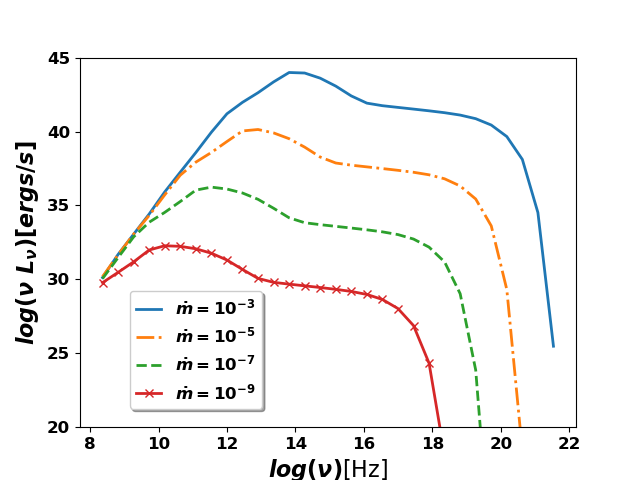}\\
   \includegraphics[height=3.5in,width=5.3in]{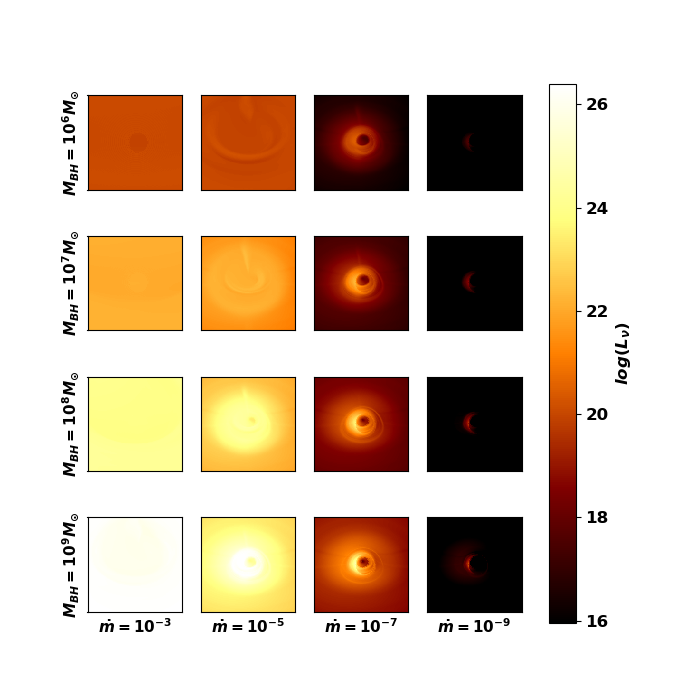}
  \caption{Spectra and intensity maps for the simulation setup {\em 21EF}.
     {\it Top left:} The total synchrotron spectra, applying the same Eddington ratio $\dot{m}=10^{-4}$, 
     but for black hole masses 
     $4\times10^6 M_{\odot}$ (blue solid), $4\times 10^7 M_{\odot}$ (yellow dot dashed), 
     $4\times10^8$ (green double dashed) and $4\times10^9 M_{\odot}$ (red crossed).
     {\it Top right:} The total synchrotron spectra, applying the same black hole mass of $5.5 \times 10^7 M_{\odot}$,
     but Eddington ratios of $10^{-3}$ (blue solid), $10^{-5}$ (yellow dot dashed), $10^{-7}$ (green dashed) 
     and $10^{-9}$ (red crossed).
     Below we show the $50~R_g$ section of intensity maps (in log luminosity scale) at 230 GHz for different black 
     hole masses and accretion rates. 
     The rows correspond to black hole masses of 
     $4\times10^6 M_{\odot}$, $4\times 10^7 M_{\odot}$, $4\times10^8$ and $4\times10^9 M_{\odot}$ 
     ({\it from top to bottom}), 
     while the columns correspond to Eddington ratios of $10^{-3}$, $10^{-5}$, $10^{-7}$ and $10^{-9}$
     ({\it from left to right}). 
     All intensity maps and spectra consider an inclination angle of $50\degree$.
  }
\label{fig:massaccspec}
\end{center}
\end{figure*}

\subsubsection{Black hole mass and accretion rate}
As mentioned above, the HARM simulation code generates outputs in gravitational units ($G=c=M=1$), and the physical scaling of quantities such as the density, pressure, temperature, magnetic field strength etc. needs to be 
obtained from the data by converting the HARM output parameters from code units to physical units. 
The mass of the black hole provides an overall physical dimension for the system. 
Also as already mentioned the HARM code provides an accretion rate in code units which needs to be converted to physical 
units with a mean accretion rate in physical units (CGS units in GRTRANs) to incorporate the gas density existing in physical systems. 
Both these quantities then allow to determine parameters such as the number density of electrons, the temperature of the 
gas and the magnetic field strengths at each point in the simulation grid. 

In order to determine the effect of black hole mass and accretion rate, we have scaled the central mass in $M_{\odot}$ units
and the accretion rate in terms of the Eddington ratio $\dot{m}$, which is the ratio of the accretion rate to the 
Eddington accretion rate, with $\dot{M}_{\rm Edd}\approx 10~L_{\rm Edd}/c^2$. 
This method of defining accretion rates relative to the Eddington rate is popular from an observational point of view, where the bolometric luminosities are 
typically expressed in terms of the the Eddington luminosities. In this investigation we are interested in systems with sub-Eddington accretion ({\it i.e.} $\dot{m}<0.01$) where the emission from the systems is not too high. It is essential for the systems to have a low luminosity in order to distinguish the various features resulting from the emission from accretion, winds, jets etc. close to the black hole.
We investigate the effect of varying masses and Eddington ratios on the synchrotron emission spectra for model {\em 21EF}. 
In Figure [\ref{fig:massaccspec}] we show a set of spectra resulting from different choices of the central mass and accretion rate.
Here, we first fixed the Eddington ratio $\dot{m}=10^{-4}$, varying the black hole mass.
We then fixed the black hole mass at $5.5 \times 10^7 M_{\odot}$ and vary the Eddington ratio. 
The black hole mass and accretion rate are chosen such that they correspond to intermediate values within the range of our interest and are somewhat comparable to Cen~A, which is a radio loud Low-Luminosity AGN (LLAGN) that is also the one closest to us. 
Furthermore, these values for the mass and accretion rate are chosen such that the spectra and intensity maps display most of the features of interest for 
all our simulation models. Finally towards the end of the paper, we also apply our models for a black hole mass and accretion rate as deduced by the EHT results for M87 for a preliminary comparison \citep{EHTC2019a}.

{We would like to note here that radiative effects may impact the dynamics of the systems with increasing accretion rate but in this work we have ignored the effects of radiative feedback in this work.}
\begin{figure*}
 \begin{center}
     \includegraphics[height=2.0in,width=2.7in]{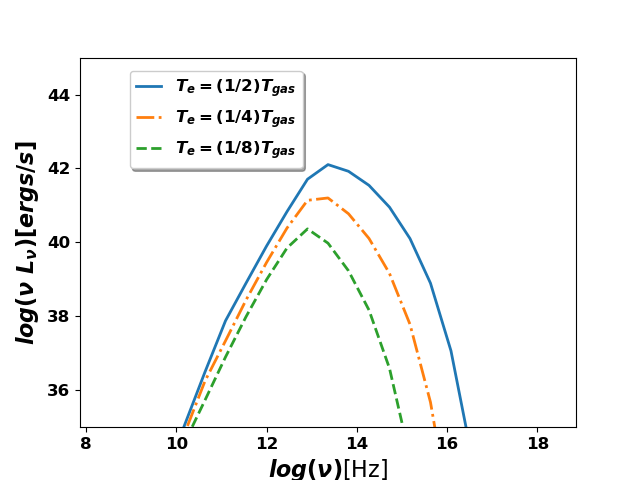}
     \includegraphics[height=2.0in,width=2.7in]{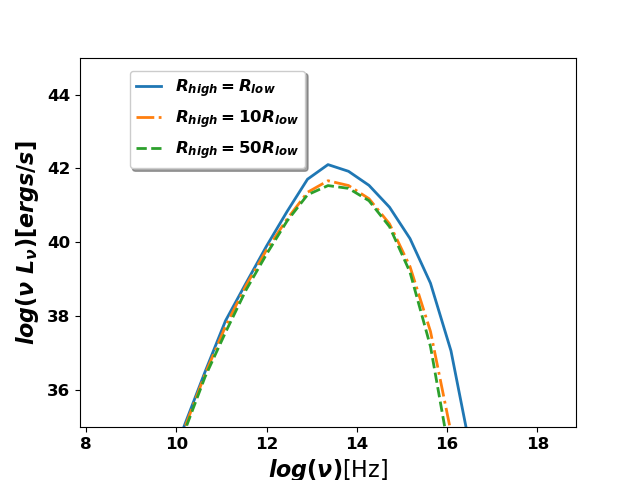}\\
     \includegraphics[height=2.3in,width=5.5in]{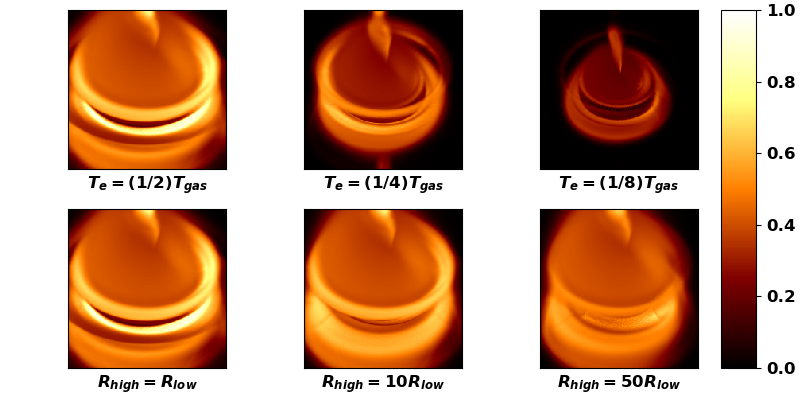}
  \caption{Spectra and intensity maps under different thermal state of the fluid for model {\em 21EF}. {\it Top left:} The thermal synchrotron spectra for electron temperature values of $T_e=(1/2)~T_{\rm gas}$ (blue), $T_e=(1/4)~T_{\rm gas}$ (yellow dot dashed) and $T_e=(1/8)~T_{\rm gas}$ (green dashed). {\it Top Right:} The thermal synchrotron spectra with $R_{\rm high}=R_{\rm low}$ (blue), $R_{\rm high}=10~R_{\rm low}$ (yellow dot dashed) and $R_{\rm high}=50~R_{\rm low}$ (green dashed) for a fixed value of $R_{\rm low}=1$. {\it Bottom:} The $50 R_g$ section of intensity maps at 230 GHz with the first row corresponding to different electron temperature and the lower row corresponds to the different choices of $R_{\rm high}$ values. For all the spectra and intensity maps, we have assumed a black hole mass of $5.5 \times 10^7 M_{\odot}$ , an Eddington ratio $\dot{m}=10^{-4}$ and an inclination angle of $50\degree$.}
\label{fig:temp}
\end{center}
\end{figure*}

\begin{figure*}
 \begin{center}
 \includegraphics[height=2.5in,width=2.5in]{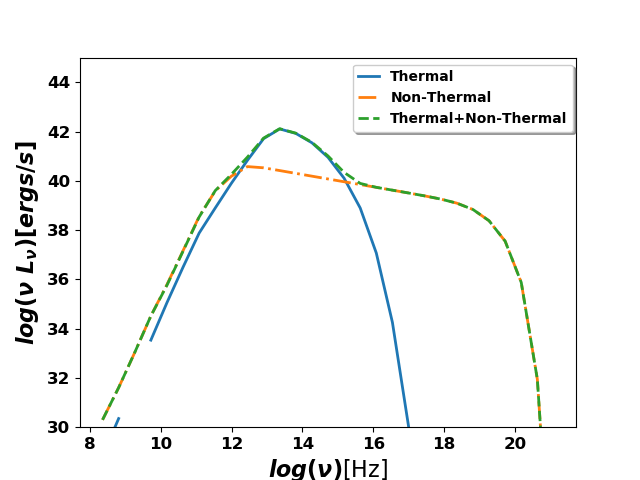}
   \includegraphics[height=2.7in,width=3.7in]{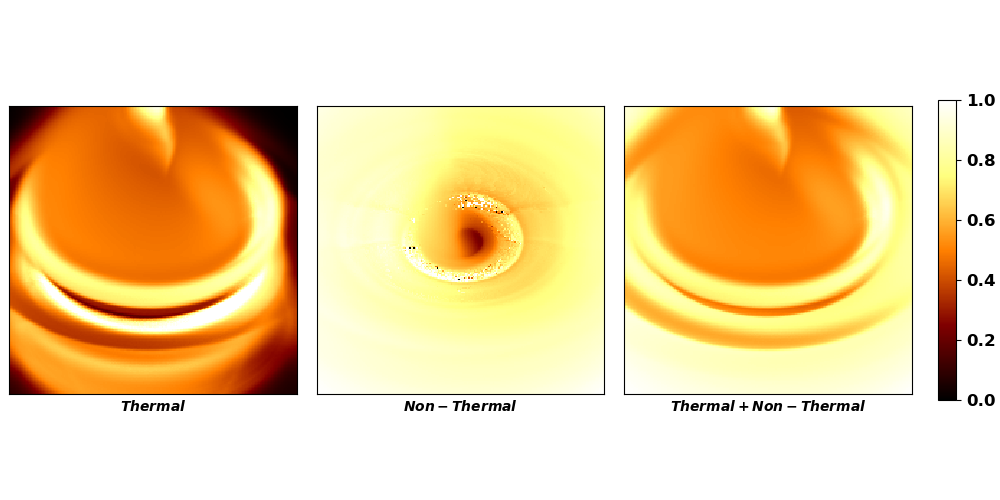}
  \caption{The spectra and intensity maps for the thermal, non-thermal and total synchrotron emissions for model {\em 21EF}. {\it Left:} The spectra showing only thermal (blue), only non-thermal electron (yellow dot dashed), and total synchrotron (green dashed) emission. {\it Right:} $50~R_g$ section of intensity maps for thermal, non-thermal and total synchtron emissions at 230 GHz respectively. The luminosity in all the frames are normalized with respect to the maximum luminosity of the total synchrotron emission frame. We have assumed a black hole mass of $5.5 \times 10^7 M_{\odot}$, a Eddington ratio $\dot{m}=10^{-4}$ and an inclination angle of $50\degree$ for all the spectra and intensity maps. The static pixels that can be seen on some of these images are numerical artifacts and do not have any physical relevance.}
\label{fig:DF}
\end{center}
\end{figure*}

\begin{figure*}
 \begin{center}
 \includegraphics[height=2.5in,width=2.5in]{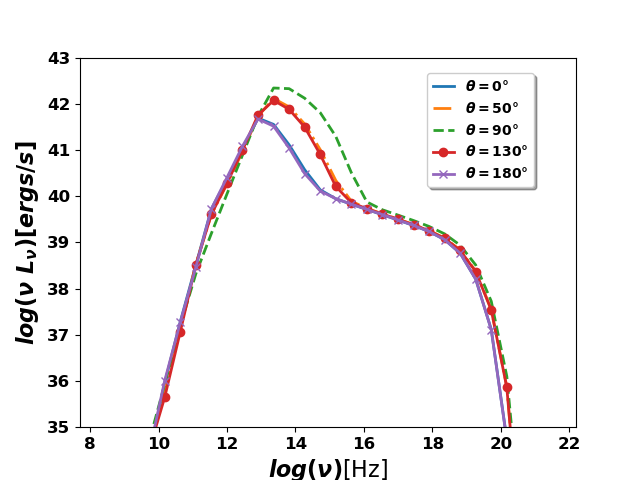}
   \includegraphics[height=2.7in,width=3.7in]{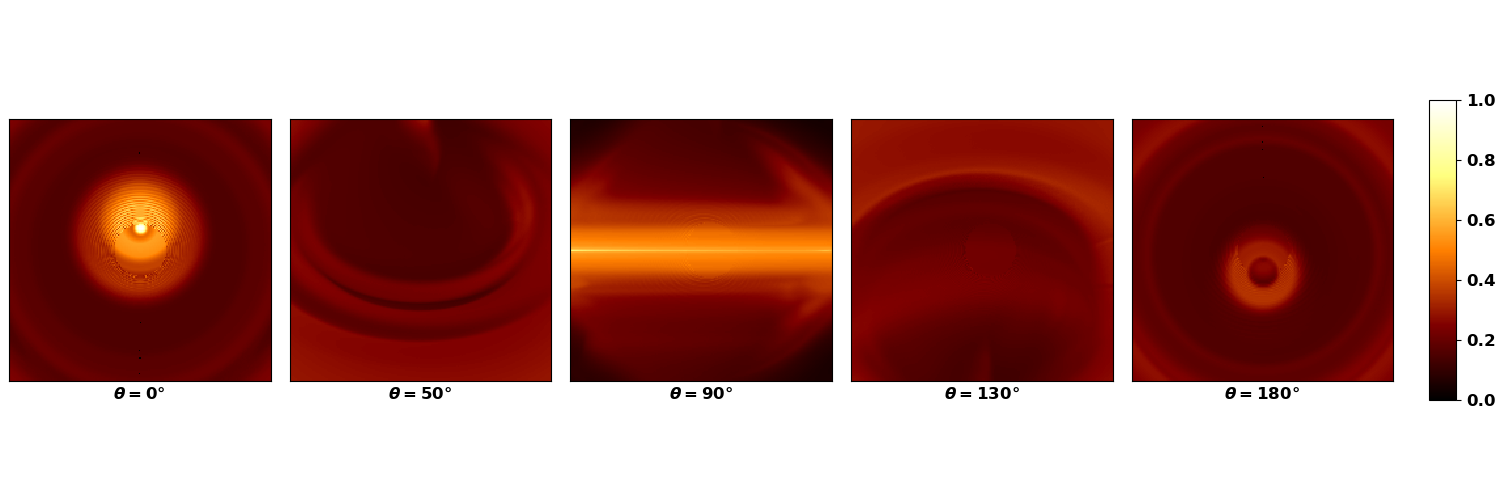}
  \caption{The spectra and intensity maps for model {\em 21EF} for a range of inclination angles. {\it Left:} The spectra corresponding to inclination angles of $0 \degree$ (blue), $50 \degree$ (yellow dot dashed), $90 \degree$ (green dashed), $130 \degree$ (red circled) and $180 \degree$ (purple crossed) with respect to the observer. {\it Right:} The $50 R_g$ section of intensity maps for the viewing angles of $0\degree$, $50\degree$, $90\degree$, $130\degree$ and $180\degree$ at 230 GHz respectively. We have considered a black hole mass of $5.5 \times 10^7 M_{\odot}$, Eddington ratio $\dot{m}=10^{-4}$ and a total (thermal+non-thermal) synchrotron emission to generate the spectra and the intensity maps.}
\label{fig:incang}
\end{center}
\end{figure*}

\begin{figure*}
 \begin{center}
 \includegraphics[height=2.5in,width=2.7in]{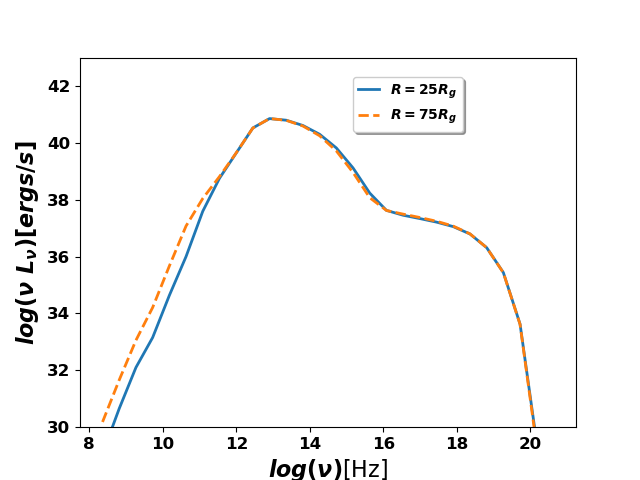}
   \includegraphics[height=2.4in,width=3.2in]{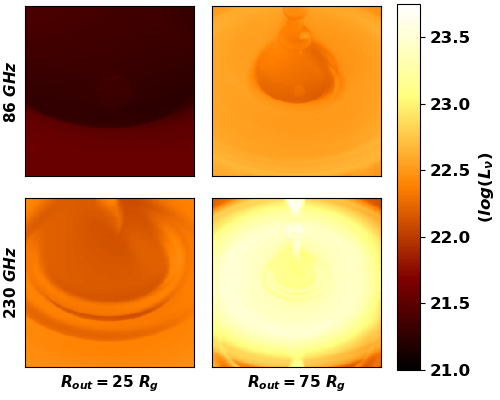}
  \caption{The spectra and intensity maps for two different physical resolutions of model {\em 21EF}. {\it Left:} The spectra for higher (blue) and lower (yellow dashed) resolutions. {\it Right:} The intensity maps for the corresponding resolutions at 86 GHz and 230 GHz respectively. For these spectra and intensity maps, we have considered a black hole mass of $5.5 \times 10^7 M_{\odot}$, an Eddington ratio of $\dot{m}=10^{-4}$ and an inclination angle of $50\degree$.}\label{fig:Rout}
\end{center}
\end{figure*}

\begin{figure*}
 \begin{center}
 \begin{subfigure}[b]{0.45\textwidth}
         \centering
        \includegraphics[height=2.5in,width=3.3in]{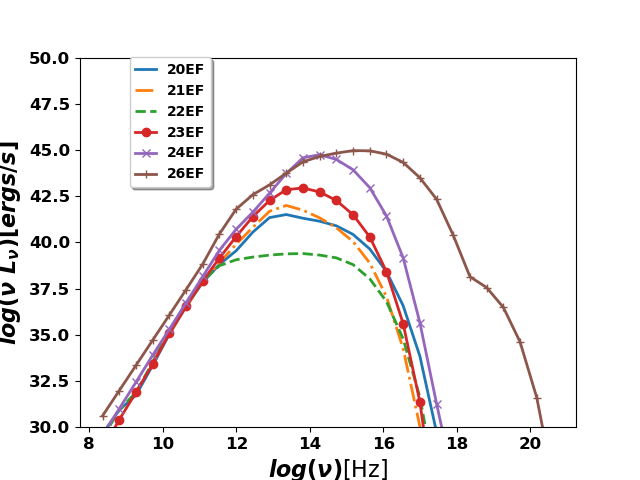}
         \caption{Thermal}
         \label{fig:subther}
     \end{subfigure}
     \hfill
     \begin{subfigure}[b]{0.45\textwidth}
         \centering
        \includegraphics[height=2.5in,width=3.3in]{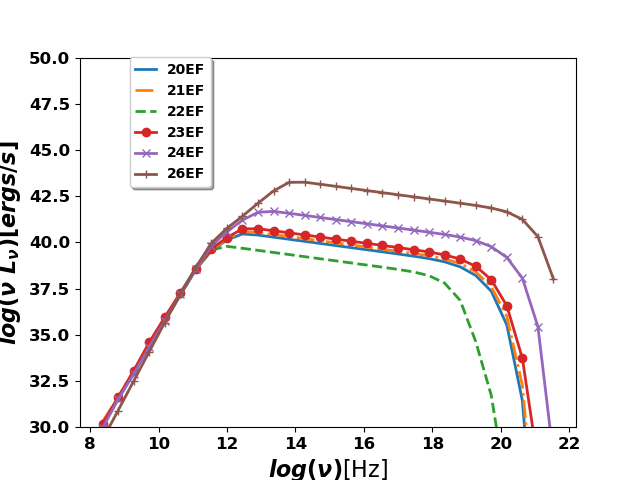}
         \caption{Non-thermal}
         \label{fig:subnon}
     \end{subfigure} \\
     \begin{subfigure}[b]{0.5\textwidth}
         \centering
        \includegraphics[height=2.5in,width=3.3in]{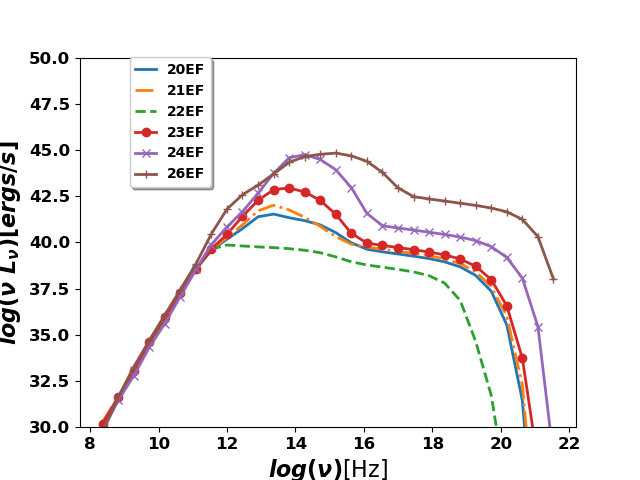}
         \caption{Thermal+Non-thermal}
         \label{fig:subtherpnon}
     \end{subfigure}
   
  \caption{ The spectra for all the simulations for thermal (\ref{fig:subther}), non-thermal (\ref{fig:subnon}) and total synchrotron emission (\ref{fig:subtherpnon}) for the model simulations {\em 20EF}(blue), {\em 21EF} (yellow dot dashed), {\em 22EF} (green dashed), {\em 23EF} (red crossed) {\em 24EF} (purple circle dashed) and {\em 26EF} (brown plus dashed). We have assumed a black hole mass of $5.5 \times 10^7 M_{\odot}$, an Eddington ratio $\dot{m}=10^{-4}$ and an inclination angle of $50\degree$}.\label{fig:allspec}
\end{center}
\end{figure*}

\subsubsection{Fluid temperature and electron distribution} 
\label{sec:temDF}
The synchrotron emission depends on the temperature of the electrons and their distribution function. 
It is thus important to determine the temperature of the electrons assuming that the gas is in plasma state.
The latter is a valid assumption since the gas temperature is high and the density being low (due to the sub-Eddington accretion 
flows considered here). 
Ideally this should come from simulations considering a two fluid plasma model in a general relativistic setup but currently such 
simulations are still quite computationally expensive.

In the current scenario, an alternate prescription is formulated which seems to replicate the plasma model to a good approximation. 
The gas temperature at each point in the grid is determined from the internal energy at those grid points assuming an ideal gas. 
Assuming a plasma consisting of only ionised hydrogen, in order to determine the electron temperature, we follow the prescription of \citet{Moscibrodzka2009, Moscibrodzka2016}, where the 
electron temperature is approximated from the following relation as
\begin{eqnarray}
    T_{e}&=&T_{\rm gas}/(1+R)\nonumber\\
    R & = & \frac{T_{\rm p}}{T_{\rm e}} = R_{\rm high}\frac{b^2}{1+b^2}+R_{\rm low}\frac{1}{1+b^2},
\label{eqn:temp}
\end{eqnarray}
with $b=\beta/\beta_{\rm crit}$, $\beta=P_{\rm gas}/P_{\rm mag}$ and $P_{\rm mag}=B^2/2$. 
The value of $\beta_{\rm crit}$ is assumed to be unity, and $R_{\rm high}$ and $R_{\rm low}$ are the temperature ratios that describe the 
electron-to-proton coupling in the weakly magnetized ({\it e.g. }discs) and strongly magnetized regions ({\it e.g.} jets), 
respectively. 

Although the total gas temperature is determined from ideal gas equations (from internal energy, density etc.), 
the choice of $R_{\rm high}$ and $R_{\rm low}$
is essential to understand the region (disc or jets) which emits predominantly.
In this work we choose $R_{\rm high}$ as a multiple of $R_{\rm low}$. 
When $R=1$, the electrons and protons on an average have equal temperature throughout the simulation box, they are hence thermally coupled. 
The electron temperature in this case is approximately half of the gas temperature. 

A ratio $R$ ({\it i.e.} $T_{\rm p}/T_{\rm e}$) greater than unity implies accretion flows where the ion temperature is greater 
than that of the electrons and the excess energy generated cannot escape efficiently. 
On the other hand a value $R< 1$ would imply that the protons are efficiently cooling down by exchanging the excess heat 
with the electrons and such systems may be radiatively efficient. 
Also increasing the value of $R_{\rm high}$ will reduce the electron temperature in the disc and hence the emission from the disc will reduce correspondingly.

As mentioned above, all GR-MHD simulations include a floor density and pressure to avoid extreme low densities. This unphysical floor value will leave an imprint on the emission model. Also the magnetisation parameter $\sigma_i$, which compares the magnetic energy to the rest mass energy of the fluid, given by
\begin{equation}
    \sigma_i =\frac{|B|^2}{4\pi \rho c^2},
\end{equation}
in general has a value of $\sigma_i<1$ in the accretion region close to the black hole. The value of $\sigma_i$ exceeds the value of 1 in the jet (
{high magnetised funnel region}) close to the poles, implying the plasma dynamics in those regions will be determined by the magnetic field. Thus any small error in the conserved energy in the simulation can lead to large errors in the energy of the fluid and thus the temperature. Thus in order to avoid the unphysical impact of floor values and the error resulting due to the large magnetic fields, we avoid regions with $\sigma_i > 1$ \citep{Chael2018}.

In addition to the thermal synchrotron emission, there can be emission from the non-thermal electrons, which can modify 
the total synchrotron spectrum. 
Such emission is important in systems where the density of the gas is really low, such as the gas in the centre of our Galaxy. 
The emissivity for a given distribution function of electrons is given as
\begin{equation}
    j=\int_{1}^{\infty}d\gamma N(\gamma)\eta \label{eqn:RT},
\end{equation}
where 
\begin{equation}
    \eta=\frac{\sqrt{3}e^2}{8\pi c}\nu_{B}sin \theta_{B}H(\nu,\theta_B)
\end{equation}
is the vacuum emissivity \citep{Melrose1980} with $e$ as the electron charge, $c$ as the speed of light, 
$\nu_B=\frac{eB}{2\pi mc}$, $H=F\left(\frac{\nu}{\nu_c}\right)$, $\nu$ is the emitted frequency, 
$\gamma$ is the electron Lorentz factor, $\nu_c=(3/2)\nu_B sin\theta_{B}\gamma^2$ and 
$F(x)=x\int_{x}^{\infty}dyK_{5/3}(y)$ is an integral of the modified Bessel function.
For a thermal distribution the distribution function in eqn.[\ref{eqn:RT}] is
\begin{equation}
    N(\gamma)=\frac{n\gamma^2\beta \exp(-\gamma/\theta_e)}{\theta_e K_2(1/\theta_e)} \label{eqn:DFTher},
\end{equation}
where $n$ is the number density of electrons and $\theta_e=\frac{kT_{\rm e}}{mc^2}$ is the dimensionless electron temperature. 
The power law distribution function is then given by
\begin{eqnarray}
    N(\gamma) = n(p-1)
                \left(\gamma_{1}^{1-p}-\gamma_{2}^{1-p}\right)^{-1} \gamma^{-p},
                  &~~&{\rm for}\,\, \gamma_1 < \gamma < \gamma_2 \nonumber\\
    0,&~~& otherwise \label{eqn:DFNT}
\end{eqnarray}
where $p$ is the index defining the power law and $\gamma_{1,2}$ are the low and high energy cut-offs, respectively. However for the jet emission, the non-thermal electron density $n$ in the above equation (also referred to as $n_{nth}$) is calculated in a slightly different way. Since the jet is magnetically dominated, the particle density and internal energy from the simulations are inaccurate due to the artificially enforced floor values. Instead of using these compromised values, the non-thermal particle density $n_{nth}$ \citep{Broderick2010} is given by
\begin{equation}
    n_{nth}=\alpha\frac{b^2}{8\pi}\frac{p-2}{p-1}(mc^2\gamma_{min})^{-1}.
\end{equation}
For our choice of $p$ and $g_{min}$, we have fixed the $\alpha=0.02$. We would however stress the fact that earlier while considering the thermal state of the matter close to the black hole, we had adopted an artificial method of ignoring regions close to the black hole with $\sigma_{i}>1$. This does not interfere with our prescription for the non-thermal electron distribution in the jets, as the non-thermal electron density in the jets does not depend on the thermal electron density. 
In this work we keep the cut-off frequencies and the power-law index fixed for all our simulations. 
It is the distribution function and the temperature distribution within the plasma, which play a vital role in determining the total synchrotron emission. In some cases we have adopted a hybrid model for the synchrotron emission {\it i.e.} where the emission from both the thermal and non-thermal emission is accounted for. For further details on the derivation, refer to \citet{Dexter2016}.

\subsubsection{Other parameters}
Other important parameters which can affect the total spectrum 
are (i) the angle at which the plane of the accretion flow is oriented with respect to the observer, 
(ii) the resolution of the image, and
(iii) the total region from which the rays are integrated. 
Also, the spin of the black hole is another important parameter which can change the photon trajectories in 
regions close to the black hole, but for data simulated by HARM, the simulation itself considers the black hole spin,
and thus a post-processing choice for the spin value is not discussed here. 
The appearance of the image also depends on (iv) the frequency at which it is observed. 
In this work we explore the entire parameter space (i)-(iv) and disentangle how the modeled spectra depends 
on these parameter choices.

\section{Results: intensity maps and spectra} 
\label{sec:Results}
Our primary aim here is to identify signatures of accretion, winds and jets from GR-MHD data through their simulated intensity maps and spectra. 
Physically, black holes with larger mass will have larger gravitational radii which implies larger angular radii subtended in the sky for a given distance (as for observations it is the $M_{BH}/D$ ratio which matters) and hence are better candidates for 
observing the outflows and winds generated in the vicinity of the black hole . 
Also, for a given telescope resolution, the most massive black holes would allow to probe accretion and jet physics at larger distances from the observer. 
In addition to the black hole mass, the physical resolution also depends on the distance from us. 
The emission is determined by the density of the gas, the magnetic field around the black hole, as 
well as the electron temperature, the electron distribution function and other parameters. 

In this section we will first consider model run {\em 21EF} as the basis for our parametric study. 
We then discuss the intensity maps and spectral properties for simulations tracing different dynamics. 
In the end we investigate the appearance of these models post-processed with the parameters from the first M87 results \citep{EHTC2019a}.  
In general we have chosen a black hole mass of $5.5 \times 10^7~M_{\odot}$ and an Eddington ratio $\dot{m}=10^{-4}$ for all 
our parametric studies. 
As already mentioned, this choice of mass and Eddington ratio corresponds to that of Cen A which is a nearby radio loud AGN,
and it allows to distinguish various features in our model simulations, therefore providing the opportunity to study their expected appearance with GRTRANS. The $R_{\rm high}$ and $R_{\rm low}$ in the temperature equation are set to unity. 

We would like to mention here that some of the images generated here display static pixels and we would like to clarify to the readers that they result only due to numerical artifacts and none of the conclusions of this work are affected by their presence.

\begin{figure*}
 \begin{center}
 \includegraphics[height=2.5in,width=3.0in]{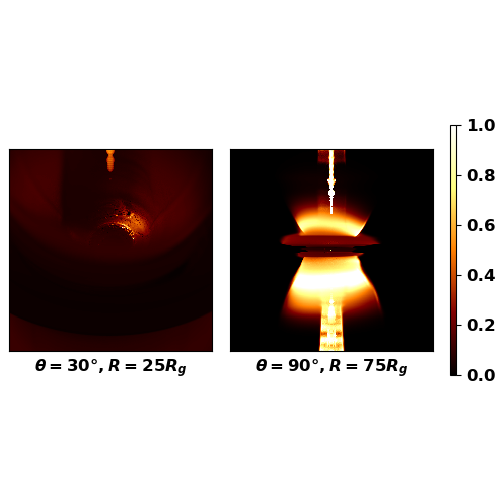}
   \includegraphics[height=2.5in,width=3.0in]{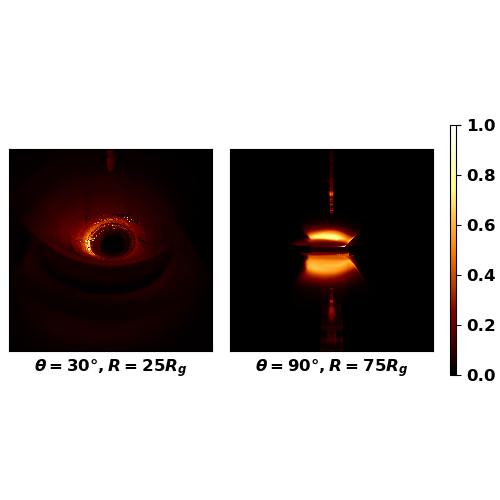}
  \caption{The intensity maps for model {\em 20EF}. The first and third intensity maps have a width $50R_g$ across at 30 degree inclination angle while the second and fourth intensity maps have a width of $150~R_g$ with an edge-on inclination with respect to the observer. The first two images correspond to 86 GHz and the last two are at 230 GHz. In these cases, we have considered a black hole mass of $5.5 \times 10^7 M_{\odot}$ and and Eddington ratio $\dot{m}=10^{-4}$. The static pixels that can be seen some of these are numerical artifacts and do not have any physical relevance.}\label{fig:Sim20_im}
\end{center}
\end{figure*}

\begin{figure*}
 \begin{center}
\includegraphics[height=6.0in,width=6.0in]{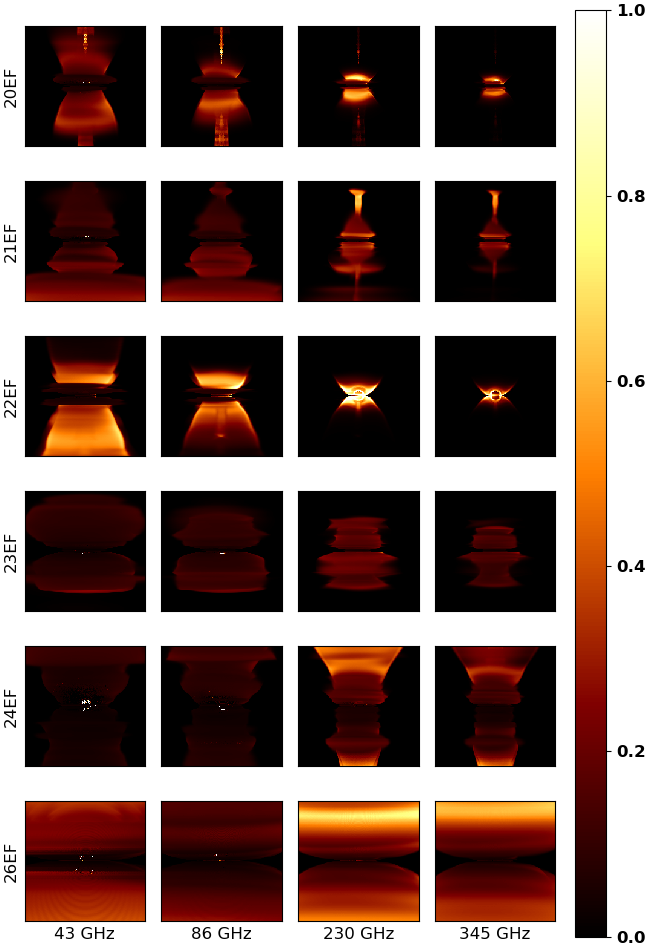}
  \caption{The intensity maps for models {\em 20EF}, {\em 21EF}, {\em 22EF}, {\em 23EF}, {\em 24EF} and {\em 26EF} ({\it rows}) for frequencies 45 GHz, 86 GHz, 230 GHz and 345 GHz ({\it columns}). In these cases, we have considered a black hole mass of $5.5 \times 10^7 M_{\odot}$ and and Eddington ratio $\dot{m}=10^{-4}$. The images are at an edge-on inclination and the width of each is $150~R_g$ across. The intensities in each of the images are normalized with the maximum intensity for each frequency band. The static pixels that can be seen some of these are numerical artifacts and do not have any physical relevance.}\label{fig:AllSim_im}
  \end{center}
\end{figure*}

  \begin{figure*}
 \begin{center}
   \includegraphics[height=2.6in,width=6.0in]{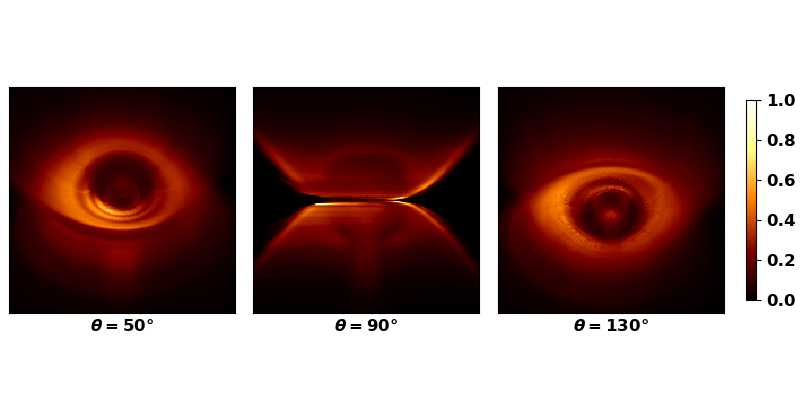}
  \caption{The intensity maps for model {\em 22EF}.These correspond to inclination angles of $50\degree$, $90\degree$ and $130\degree$ respectively. The width of each of the images here is $50~R_g$ across. The black hole mass and Eddington ratio considered in this case are $5.5 \times 10^7 M_{\odot}$ and $\dot{m}=10^{-4}$ respectively}\label{fig:Sim22_im}
\end{center}
\end{figure*}


\subsection{Impact of model parameters on intensity maps and spectra}

Before we analyze the features obtained by ray-tracing all our models, we first want to
demonstrate how a variation of the leading parameter choices may change the appearance of 
our spectra and intensity maps.

\subsubsection{Impact of black hole mass and accretion rates}

In the top panels of Figure~[\ref{fig:massaccspec}] we show the spectra for model {\em 21EF} to understand the impact
of the black hole mass and the accretion rate on the overall synchrotron spectra. 
The spectra in both the panels display two peaks, the first corresponds to the thermal synchrotron emission while the second corresponds to 
non-thermal synchrotron emission. 
These spectral features are specific to the particular simulation under consideration (in this case 
model {\em 21EF}) and we will discuss later how features change for different simulations. 
The top left panel of the figure is generated by applying a fixed Eddington ratio of $\dot{m}=10^{-4}$
and varying the black hole mass between $\approx10^6M_{\odot}$ (corresponding to the mass of Sgr A*)
to $\approx 10^9M_{\odot}$ (corresponding to the black hole mass of M87). 
Since the black hole mass determines the Eddington accretion rate, the accretion rate in physical scales thus varies
between $\sim 8.9\times 10^{-7}M_{\odot}y^{-1}$ to $\sim 8.9\times 10^{-4}M_{\odot}y^{-1}$, respectively. In addition to the higher accretion rates, the mass of the black hole converts all quantities to physical 
units (which were initially in units of $G=c=M=1$ units) in the system, 
implying higher luminosities for higher black hole masses at all frequencies investigated (same figure).

On the other hand, by fixing the black hole mass but varying the Eddington ratio implies varying the accretion rate. 
We fix the black hole mass ($5.5\times 10^7$) and vary the Eddington ratio between $10^{-3}$ to $10^{-9}$. 
The range of Eddington ratios is chosen such that we can explore a very large range of Eddington ratios for sub-Eddington accretion rates.
A higher accretion rate then implies higher densities and velocities in the vicinity of the black hole and
thus a higher thermal state for the electrons. 
This finally results in higher luminosities at higher frequencies for higher accretion rates as shown in 
the spectra of Figure~[\ref{fig:massaccspec}, upper right].
    
In the bottom panel of the same figure are shown the $50~R_g$ width intensity maps at 230 GHz for the same simulation 
model {\em 21EF}. It displays the impact of varying the black hole mass and the Eddington ratios on the overall luminosity at 230 GHz. A log scale is chosen to display the emission features because it enables to distinguish the various emission features over the scale of black hole masses and Eddington ratios.
We observe in this image that for a given Eddington ratio, the overall luminosity increases for a higher black hole mass 
while for a given black hole mass, the visibility of the various components of the image changes by varying the 
Eddington ratios.
A higher Eddington ratio of $\dot{m}=10^{-3}$ implies very high electron densities which results in a strong emission from the outer winds and jets which completely 
blocks the visibility of innermost regions. 
This kind of effect will be more pronounced in systems with very strong disc winds as we will later see in case of simulation model {\em 24EF} and {\em 26EF}.
On the other hand, for a low Eddington ratio of $\dot{m}=10^{-9}$ which implies very low density for a given black hole mass, the emission from the the innermost regions is visible where the density is relatively high compared to the surrounding regions. The radiation thus escaping from the innermost regions are significantly affected by the strong gravity of the black hole and are thus beamed, boosted and lensed to produce the ring-like appearance around the black hole.  

Overall, for such kind of systems, with a higher black hole mass of $\simeq 10^9 M_{\odot}$ and a lower Eddington ratio of
$\simeq 10^{-9}$, we would be able to identify a dark region within the photon ring (we will from now on refer to
it as the black hole shadow). 
On the other hand for an intermediate black hole mass of $\simeq 10^7 M_{\odot}$ with an Eddington ratio $\simeq 10^{-5}$, which implies relatively higher density such that the emission from the innermost region is blocked by the emission from the disc wind and jet. The density is still not too high and thus allows us to identify features such as disc winds and jets. One  can clearly observe the jet from the centre and the surrounding disc wind in the emission map at 230 GHz. We would like to emphasize here that in general jets and wind structures are clearly visible in systems with moderately high accretion rates or higher densities while systems with lower accretion rates display the ring-like features which can either be from the disc or the lensed emissions from disc winds and jets. In this work we will subsequently concentrate on the intermediate case of a black hole mass $\simeq 10^7$ and an Eddington ratio of $\simeq 10^{-4}$) for our further analysis, as we can identify features of accretion, disc wind and jet.

\begin{figure*}
 \begin{center}
 \includegraphics[height=3.5in,width=5.3in]{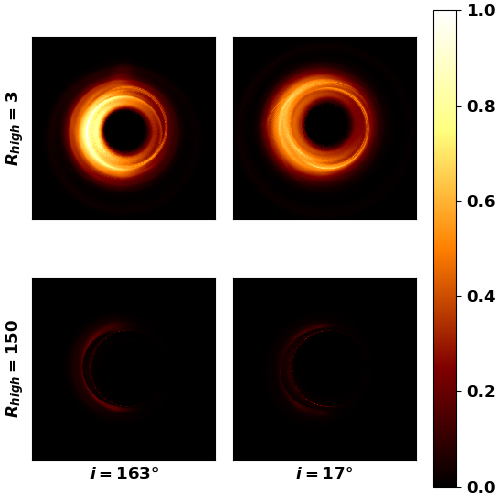}
  \caption{Intensity maps of model {\em 21EF} for M87 like black hole mass of 
  $M_{\text{BH}}=6.5 \times 10^{9}M_{\odot}$, $\dot{m}=2 \times 10^{-5}$, $D=16.8 \text{Mpc}$ and $T_{\rm p}=3 T_{\rm e}$. 
  The rows correspond to $R_{\rm high}$ values of 3 and 150 respectively and the columns correspond to inclination angles of 163 and 17 degrees with respect to the observer respectively. The fluxes are normalized with the maximum flux in the upper left panel. The width if the images are $24~R_g$. The upper panels correspond to an integrated flux of about 2 Jy and the lower panels correspond to integrated flux of about 1 Jy.}\label{fig:M87ring}
\end{center}
\end{figure*}
  
  \begin{figure*}
 \begin{center}
 \includegraphics[height=3.0in,width=8.5in]{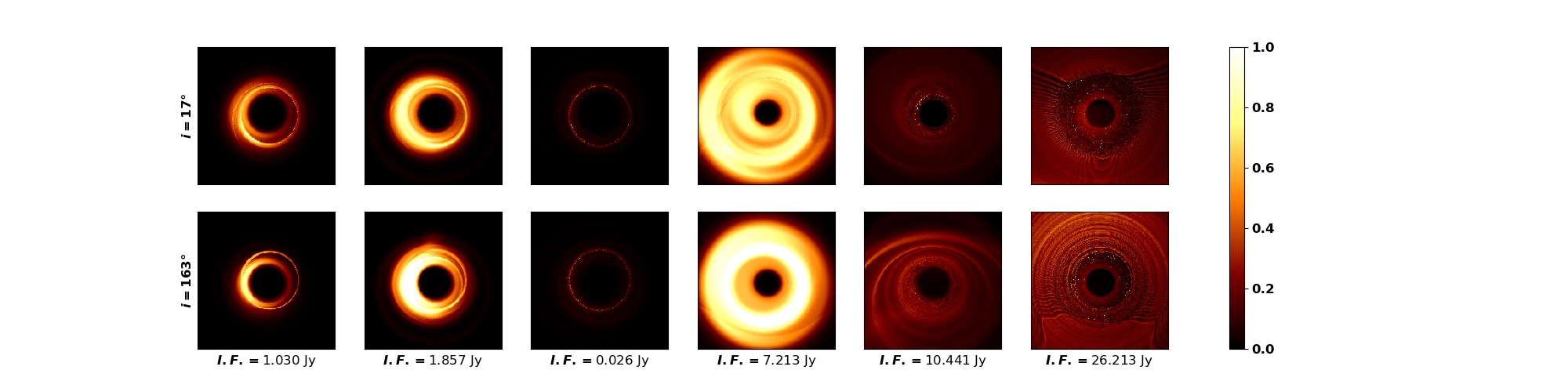}
  \caption{Intensity maps for models {\em 20EF}, {\em 22EF}, {\em 23EF}, {\em 24EF} and {\em 26EF} with the x-labels displaying the integrated flux are shown for each columns respectively for M87 like black hole mass of $M_{\text{BH}}=6.5 \times 10^{9}M_{\odot}$, $\dot{m}=2 \times 10^{-5}$, $D=16.9 Mpc$, $T_p=3T_e$ and $R_{\rm high}=150$. The width of the intensity maps in each case is $24~R_g$. The rows correspond to inclination angles of $17\degree$ and $163\degree$ respectively with respect to the observer. The intensity maps in each column are normalized corresponding to the maximum flux corresponding to that column. The static pixels in some of these images are numerical artifacts and do not have any physical relevance.}\label{fig:M87Allsim}
\end{center}
\end{figure*}  

\subsubsection{Impact of electron temperature and distribution function}
Synchrotron emission depends on the number density of electrons, temperature of the electrons, the strength 
of the magnetic field and the distribution function of electrons at each grid point.  
The primary difference between the thermal and non-thermal synchrotron emission is that the non-thermal electron distribution function
does not depend on the temperature of the electrons and is primarily determined by a power-law (equation[\ref{eqn:DFNT}]). 
Non-thermal electrons could be generated due to turbulence, the presence of magnetic fields, winds and jets.

In order to understand the effect of thermal variations, we thus investigate the thermal synchrotron emission only. 
Figure~[\ref{fig:temp}] and Figure~[\ref{fig:DF}] summarise the effect on the intensity maps and spectra for 
different choices for the thermal state and the distribution function of electrons. 

We have varied the emission from different regions of the model by changing the temperature of the electrons using eqn.~[\ref{eqn:temp}]. 
A higher $R_{\rm low}$ decreases the electron temperatures in the jet (thus lower emission from the jets and winds ) while a higher
width of $R_{\rm high}$ implies lower electron temperatures in the disc and hence less radiation from the disc. 
We begin by changing the average electron temperature i.e. we fix $R_{\rm high}=R_{\rm low}$ 
to values of 1, 3 and 7, such that the average electron temperature varies as $T_{\rm e} = (1/2),(1/4),(1/8)~T_{\rm gas}$. 
As can be seen in the upper left panel of Figure~[\ref{fig:temp}], the thermal synchrotron spectrum shifts to lower luminosities implying that the radiative efficiency decreases with the decreasing average electron temperature. 
Also, the middle panel of the same figure shows that at 230 GHz, a lower electron temperatures implies a lower overall 
total luminosity.
We also see the emission from the the jet base distinctly from the rest of the emission from the disc. 

For the simulation model {\em 21EF}, increasing the $R_{\rm high}$ values to 10 and 50 times $R_{\rm low}$ 
does not change the spectra or the intensity maps much. 
Note that increasing $R_{\rm high}$ implies decreasing the electron temperature selectively from the disc. 
Since we don't see a significant difference in the brightness of the image by increasing $R_{\rm high}$, we can infer that the thermal emission is primarily from the disc wind (which has a strong magnetic field  compared to the disc).
  
In Figure~[\ref{fig:DF}], we compare the thermal and non-thermal emission spectrum. The non-thermal synchrotron emission spectra generally extend up to higher frequencies. 
As mentioned earlier the peak of the total spectrum is determined by the thermal emission, whereas the low frequency and high frequency tails are dominated by the power law synchrotron component for this particular simulation model.

\subsubsection{Impact of other system parameters} 
We find a small change in the total emission spectra for different inclination angles. 
As can be seen in the left panel of Figure~[\ref{fig:incang}], the emission peaks for an 
edge-on view (inclination angle of $90\degree$ with respect to the observer),
while it is lowest at face-on view ($0\degree$ and $180\degree$).
This results because for an edge-on view we receive emission from both the upper and lower regions of the disc,
while in the case of a face-on view, the emission is received from only the region facing towards the observer. 
We note, however, that a closer look into the spectrum shows that at around 230 GHz, the face-on emission 
seems to increase compared to the rest of the inclination angles. 
This is because the energy of the photons from the jet have energies corresponding to that frequency band.
The intensity maps at $0\degree$ and $180\degree$ display a variation in the brightness of the jet which implies a brighter front jet compared to the backward jet. 
A large scale image of the same simulation from the edge-on view also shows that the front jet is brighter (Emission map corresponding to {\em Sim21} at 230 GHz in Figure{\ref{fig:AllSim_im}}) which implies a higher velocity in the forward jet compared to the backward jet.
    
The total flux also changes depending on the size of the region for which ray-tracing is done (Figure~[\ref{fig:Rout}]). From an observation point of view a large scale mapping corresponds to a lower resolution while a small scale corresponds to a high resolution imaging.
For a larger scale ($150~R_g$ across i.e. a radius of $75~R_g$ from the centre of the black hole), 
a larger area of the source is viewed.
Hence the total flux contribution from the surrounding area enhances the overall flux at a given frequency,
compared to the emission map in case of small scale ($50 R_g$ across i.e. $25 R_g$ from the center). This in principle  is a physical effect that depending on the region and the matter within the field of view, the spectra may appear differently. While the EHT will typically assume a compact source in its analysis, it is in principle conceivable to have extended sources for more general VLBI-approaches.
Also, as can be seen from the intensity maps in the same figure, the fluxes  at 230 GHz are higher.
Here, we probe deeper into the regions around the black hole (i.e. these emissions are closer to the black hole) compared to 86 GHz where the  synchrotron self-absorption is stronger at lower frequencies. The typical flux at lower frequencies follows the relation $S(\nu)\propto\nu^{(5/2)}$.

\subsection{Analysing different simulation maps}
We finally post-process all the dynamical models summarized in Table~[\ref{tab:para_dynamics}] with GRTRANS 
in order to understand how the synchrotron emission differs for the different simulations and the configurations they represent.
Here we have applied the same post-processing parameters as above. 
The black hole mass and Eddington ratios for all the spectra are fixed at $5.5 \times 10^7$ and $\dot{m}=10^{-4}$. 
The mean temperature of electrons is fixed as $T_e=(1/2)T_{\rm gas}$ ({\it }i.e. $R_{\rm high}=R_{\rm low}=1$).
We first compare the thermal, the non-thermal and the total synchrotron spectra obtained from these simulations. 

In Figure~[\ref{fig:allspec}] we compare the thermal (left top panel), non-thermal
(right top panel) and the total (bottom panel) synchrotron emission spectrum for each of the simulations 
for the same set of parameters. 

Model simulations {\em 24EF} and {\em 26EF} have the highest luminosities both in the thermal and non-thermal components while 
simulation {\em 22EF} has the lowest integrated luminosity. 
Simulations {\em 24EF} and {\em 26EF} that consider Kerr black holes also have low plasma-$\beta$, implying stronger
magnetic pressure which generates strong winds and jets, in contrast to simulation {\em 22EF} considering a 
Schwarzschild black hole that is devoid of strong winds and BZ jets (see Section [\ref{sec:DM}]).
The thermal synchrotron emission from model {\em 26EF} extends to frequencies up to the X-ray regime.
All our models have thermal synchrotron spectra similar to the advection dominated accretion flow (ADAF) models (for reference see \citealt{Bandyopadhyay2019,Nemmen2014}).
It is interesting to note that although our model setup corresponds to  a very different disc structure compared to the initial torus model
that is usually assumed in the literature, the resulting thermal synchrotron spectral shape are very similar to our models.
This implies that models with a Keplerian disc  but with the disc winds and jets can also lead to similar synchrotron peaks, which of course is essential when it comes to interpreting the observed spectra. We would however like to emphasize here that models with stronger magnetic field show a shift in the peak emission to higher frequencies such as {\em 26EF}.

From the upper right panel of Figure.[\ref{fig:allspec}], we notice that the shape of non-thermal spectra for all the simulations are very similar. We know from our dynamical modeling that the disc structure for all our model simulations are similar. In this plot we see that the non-thermal spectra for models {\em 20EF}, {\em 21EF} and {\em 23EF} overlap which clearly suggests that the non-thermal emission in these systems originate in the disc. Models {\em 24EF} and {\em 26EF} display higher luminosity displaying the contribution of the non-thermal electrons from the jets. Model {\em 22EF} on the other hand shows a lower luminosity compared to others. This model being the Schwarzschild case where we know that the ISCO (innermost stable circular orbit) is much larger than the Kerr case, implying the the highest energy of emissions in the non-thermal electrons originate from regions very close to the black hole. This feature of non-thermal emission from the disc is another interesting feature of our model which differs from the other simulations in literature where most of the non-thermal emission originate in the jet. The primary reason behind this is the presence of large scale magnetic fields in contrast to the MAD simulations where the strong magnetic fields exist mostly in the jet region. At frequencies of our interest, we notice that the total spectra is dominated by the thermal synchrotron emission. Hence for understanding the emission features, we will consider the thermal synchrotron emission only.

For the sake of a better understanding, we use model {\em 20EF} to display both its large scale 
and small scale intensity maps at 86 GHz and 230 GHz, respectively (Figure.[\ref{fig:Sim20_im}]). 
We have then included the large scale intensity maps at 43 GHz, 86GHz, 230 GHz and 345 GHz for all  of the other simulations with an edge on inclination (Figure.[\ref{fig:AllSim_im}]).  These frequencies employed here correspond to those typically used in observation with the VLBI technique. In addition for simulation {\em 22EF} we show a small scale intensity map at three inclination angles as can be seen in Figure [\ref{fig:Sim22_im}], which 
displays interesting features such as the emission from the infalling matter which leaves a signatures in the beamed
emission and can  only be distinguished at smaller scales ($50 R_g$).

In Figure~[\ref{fig:Sim20_im}], we clearly recognize the BZ jet base originating very close to the black hole at 86 GHz. At 230 GHz, the black hole shadow can be recognised when observed at the $\theta=30\degree$ while the large scale image with a edge on view displays the collimated spine jet for both 86 GHz and 230 GHz. The edge-on view at both 86 GHz and 230 GHz clearly show the signature of strong disc wind. The disc is however not visible with thermal synchrotron emission because of synchrotron self absorption in the high density disc.

In Figure.[\ref{fig:AllSim_im}] we can see a $150~R_g$ edge-on view of all our simulation models at different frequencies. At the lowest frequency [i.e.] 45 GHz the signature of the disc winds are most clearly visible while at the highest frequency of 345 GHz, we can see the emission signatures from the innermost regions of the AGNs. Models {\em 21EF} in addition also displays the signature of the BZ jet at 345 GHz.  
The spine jet and strong disc wind is brightest at the the lowest frequencies while the disc wind base is visible in the highest frequency for {\em 20EF}. Model {\em 21EF} also has a disc wind which extends to very large scales which is visible at frequencies of 43 GHz and 86 GHz. Model {\em 22EF} has large scale outflows visible at low frequencies and the base of the flow is visible at high frequencies. Being  a Schwarzschild black hole it does not allow for the formation of a BZ jet. There is also no signature of a BZ jet for model {\em 23EF} in any frequency, but a disc wind is clearly visible at all frequencies. This is because of the high value of the floor density which inhibits the generation of a BZ jet even though the spin of the black Hole is not 0. Models {\em 24EF} and {\em 26EF} have very strong wind signatures due to the very high magnetic field and wind speed and thus the emission is high at all frequencies. The emission from the disc wind for model {\em 26EF} is too high to make the innermost BZ jet visible from an edge on view.

The dynamical model {\em 22EF} considers the special case of a Schwarzschild black hole as mentioned, however with otherwise the same
parameters and initial conditions as model {\em 21EF} (refer Table~[\ref{tab:para_dynamics}]).
For the chosen parameters the simulation does not result in very large-scale winds or jets and thus we here extract the information from
it on smaller scales.
Figure~[\ref{fig:Sim22_im}] shows the $50~R_g$ intensity maps at 230 GHz for inclination angles of $50\degree$, $90\degree$ and $130\degree$. 
From all these intensity maps it is clearly evident that a BZ jet is absent. 
The enhanced brightness at the south-west direction of the emission map at $50\degree$ inclination and on the north 
of the emission map at 
$130\degree$ inclination could be interpreted as a signature of the infall of material close the black hole.
This infall in fact reaches relativistic speeds, potentially leading to strong Doppler boosting along with the emission from the 
innermost regions of the disc.
Note that the difference between the usual boosting of jet emission (which is boosted towards the observer due to the jet motion),
and the boosting of the infalling gas on the {\em far} side of the BH moving towards the observer. Thus the emission of this outflow is boosted or beamed, respectively, while the emission from the infalling gas on the near
side of the BH is moving away from the observer and is de-boosted.

\subsection{Application of M87-like parameter values for model simulations}
The first image of the lensed ring around the black 
hole of M87 with the Event Horizon telescope \citep{EHTC2019a,EHTC2019b,EHTC2019c,EHTC2019d,EHTC2019e,EHTC2019f} has been a significant breakthrough for the imaging of supermassive black holes.
For the sake of completeness we would like to test if our simulations produce ring like features for M87 like parameters. This also allows to test how our simulations appear under a different set of physical conditions.  
We thus consider a black hole mass of $M_{\text{BH}}=6.5 \times 10^{9}$, an Eddington ratio of 
$\dot{m}=2 \times 10^{-5}$, a distance of 16.9~Mpc, temperature partition as $T_{\rm i}=3T_{\rm e}$ and inclination angles of $17\degree$ and $163\degree$, as deduced from  the images of the 
black hole shadow by the EHT collaboration\citep{EHTC2019a,EHTC2019e}. Here for this analysis, we will consider thermal synchrotron emission only. 

For these parameter values, we test our model {\em 21EF} for $R_{high}$ values of 3 and 150 \citep{EHTC2019e} (Figure~[\ref{fig:M87ring}]). We can see an axisymmetric ring like feature for both inclination angles but the fluxes are is high for $R_{high}=3$. Since a high value of $R_{high}$ implies a lower electron temperature in the disc, a reduction in flux due to increase in $R_{high}$ thus implies that in this case most of the emission signature is due to the emission from the disc.

We then test all our simulations for the same set of parameters. Since we find that the fluxes are really low for large $R_{high}$ for some of our simulations, the emission maps shown  here in Figure. [\ref{fig:M87Allsim}] correspond to $R_{high}=3$. The images in each column (model) of the figure are normalized by the maximum intensity in corresponding to that column (model). We found that models {\em 20EF}, {\em 21EF} and {\em 22EF} show this characteristic of flux reduction, implying that the emission signatures in these cases are dominated by thermal electrons in the disc. On the other hand models {\em 23EF}, {\em 24EF} and {\em 26EF} do not display much of a change with increasing  $R_{high}$ implying that the thermal synchrotron emission in these cases is mostly dominated by the electrons in the winds and jets.

Ideally the best way to deduce the best fit flux model is by comparing the generated spectra from the simulations to the observed data but as for an exact comparison, we would need to further process our data for values such as the jet position angle, convoluting our results with $20\mu as$ FWHM Gaussian beam width etc., which is beyond the scope of this work and also not our primary motivation here. We still observe that the integrated flux from our models {\em 20EF} and {\em 21EF} display an integrated flux close to 1Jy as observed for M87 at 230 GHz. These models clearly also display the asymmetry in the ring brightness. Model {\em 20EF} in addition to the photon ring displays a brighter inner ring at both the inclination angles which indicates a signature of the disc wind base. We would like to emphasize here that our Keplerian disc model with disc winds like {\em 20EF} and {\em 21EF} show similar ring like features with similar flux values as that obtained from torus like models, is an indication that there could be alternate models which can match the observed data.   

Model {\em 22EF} shows only a ring like feature which is very faint. The thick bright emission region around the black hole shadow of model {\em 23EF} is a clear signature of the disc wind.

Models {\em 24EF} and {\em 26EF} seem to display almost equal brightness around the black hole shadow which is a clear signature of the emission from the strong disc winds in these cases. In addition we would like to point out that the 
{not-so-smooth} pixel brightness especially visible for model {\em 26EF} is due the the prescription adopted to avoid regions with $\sigma_i>1$. Although this method is generally applied to systems where the magnetic field strengths are strong near the jet axis where also there are unphysical density because of the floor models incorporated but in cases like mode {\em 26EF} this method leads to errors because of the presence of the strong magnetic field extending beyond the disc region. This effect is not there for other physical parameters such as changing the mass an accretion rate (We already saw that this effect was not present when we had adopted a lower black hole mass of $5.5\times 10^7 M_{\odot}$).


\section{Conclusions}
\label{sec:Conclusion}
In this work we have investigated the various factors which affect the radiative appearance that emerges from the accretion disc, 
the disc wind and the jets that are hosted by supermassive black holes.
These different emission features are central for the purpose of providing predictions for future VLBI observations.

Our approach was the following:
We have first simulated GR-MHD jet launching models with rHARM \citep{Vourellis2019}, in particular 
considering resistive MHD that allows for the mass loading of a disc wind by an accretion disc in Keplerian rotation.
We obtained six differently parametrized black hole-disc systems that have shown different (relative) mass loading for the
disc winds and jet flows.

In a second step we then went on and applied a post-processing relativistic radiative transfer using GRTRANS \citep{Dexter2016}
to obtain a series of intensity maps ans spectra.
For that we have scaled the simulation results (that were obtained in code units) applying different astrophysical parameters 
such as black hole mass and accretion rate.
In addition, further assumptions on the composition and the energetics of the plasma particles had to be made.

We were thus able to investigate the impact of various physical systemic parameters on the synchrotron emission.
We provided, discussed and compared emission maps and emission spectra for a wide parameter range.
We also provide, as a benchmark of our approach, results for M87-like parameter values.

In the following we summarize our results in detail.

\begin{itemize}
\item[(1)] We study the appearance of the black hole environment starting from a thin disc model, including the accretion disc, winds and jets, varying black hole mass, accretion rate, spin, inclination angle, disc parameters and observed frequency. When we adopt M87-like parameters, we show that we can reproduce a ring-like feature (similar to the observation from the EHT) for some of our simulations. The latter suggests that such thin disc models are thus likely to be consistent with the observed results.

    \item[(2)] A higher black hole mass and a high accretion rate enhances the overall synchrotron emission for a given simulation model. 
    For a model with a moderately prominent disc, wind and jet system (as in the case of simulation {\em 21EF}),
    the choice of the black hole mass primarily aids in converting code units to physical units. 
    This affects the overall energetic by contributing to the Eddington rate, which affects electron density and energy which then affects the integrated luminosity of the system.
    The Eddington ratio on the other hand determines the visibility of structures
    (whether prominent black hole shadow or prominent jets and winds) at any given frequency by altering the density of the system.
    
    \item[(3)] In our models, the disc dominates the non-thermal synchrotron emission spectra for models with moderate magnetic filed strength while the disc winds and jets
     become visible via the thermal synchrotron emission at 230 GHz for system parameters of $M_{\rm BH}=5.5\times 10^7$ and $\dot{m}=10^{-4}$. For models with high magnetic fields strength, the non-thermal spectra is dominated by the emission from the jets.
     A higher electron temperature in all system components leads to an overall higher emissivity. 
     Changing the relative temperature of the disc is not significant when the emission is dominated by the winds and jets.
    
    \item[(4)] The emission maps of all our simulations show the presence of a disc wind.
     A BZ jet is not generated for models with low black hole spin ({\em 22EF}) or low jet density ({\em 23EF} with a very high floor density).
     The visibility of the BZ jet at any frequency also depends on the relative strength of emission from the winds and jets.
     Strong disc winds can block the visibility of the innermost jets for an edge on inclination (as in the case of simulation {\em 24EF} and {\em 26EF}). The presence of the innermost BZ jet is visible in almost a face-on inclination.
    
    \item[(5)] Applying typical M87-like parameters as suggested by \citet{EHTC2019a,EHTC2019e}, 
    we can clearly observe the ring-like features for our models with moderate magnetic field strength
    (that are {\em 20EF}, {\em 21EF} and {\em 22EF}). 
    The integrated flux is almost of the order of 1 Jy (as expected for M87 at 230 GHz) for the models of moderate 
    mass fluxes and magnetic field strength, {\em 20EF}and {\em 21EF}
    whereas orders of magnitude lower for the Schwarzschild case (model {\em 22EF}).
    
    \item[(6)] Models with strong wind emission tend to smear out the photon ring emission even for the choice of M87-like parameters 
    where we also observe a high integrated flux but the shadow region is still visible. 
    
\end{itemize}    

In summary, with our results we were able to
highlight the significance of the physical parameters that impact the emission from the core of AGN with winds and jets.
Given that the accretion disc is similar for all the simulations, we can state that the difference in features that 
we observe in the intensity maps for the same set of initial parameters (including black hole mass and accretion rate) are
mostly due to the presence or absence of winds and jets. 
This is in particular also visible when we adopted M87-like parameters for our disc-wind-jet system around the black hole especially the ones with higher magnetic field strength or the one with strong disc wind.

Since the first imaging of the M87 black hole shadow, there has been a plethora of investigations in order to understand the emission from the innermost regions of the accretion flow.
As an example on similar lines of research a very recent study by \citet{Bronzwaer2021} investigated the visibility of the black hole shadow and made a comparison with the image shadow considering strong and weak emissions from the inner part of the accretion disc, while our study on the other hand broadly discusses and describes the conditions under which wind and jet features are visible in an accreting system. In the context of variable systems, \citet{Jeter20} explored how to differentiate between the disc and jet systems in the presence of hot spots. \citet{Ripperda20} investigated magnetic reconnection and hot spot formation in black hole accretion discs, which can be particularly relevant for time-dependent observations. \citet{Vincent21} pursued further modeling of the observed ring in M87, considering also deviations from the Kerr metric. Also \citet{Nampalliwar20} explored how the topology of the horizon can be probed from VLBI observations in order to probe deviations from General Relativity. 

We conclude by stating again that the main aim of our work was
to understand both the large-scale and small-scale emissions for a range of frequencies and the general spectral behaviour of systems 
with strong disc winds and jets.
With the ng-EHT and the other VLBI facilities, we will be able to observe larger numbers of sources and thus various models of accretion and emission will be necessary both to derive the expected images and to interpret the observations in the future.

\section*{Acknowledgements}
All GR-MHD simulations were performed on the ISAAC cluster of the Max Planck Institute for Astronomy.
C.F. and C.V. are grateful to Scott Noble for the possibility to use the original HARM3D code for further development.

BB thanks funding via Fondecyt Postdoctorado (project code 3190366).

BB and DRGS thank for funding via the  Millenium Nucleus NCN19$\_$058 (TITANs), ANID PIA ACT172033 and the Chilean BASAL Centro
de Excelencia en Astrofisica y Tecnologias Afines (CATA) grant PFB-06/2007. 

The authors thank Jason Dexter for his suggestions related to GRTRANS. BB and DRGS thank for stimulating discussions with Neil Nagar, Venkatessh Ramakrishnan, Javier Lagunas and Javier Pedreros on related topics.


\section*{Data Availability}
Data available on request. The data underlying this article will be shared on reasonable request to the corresponding author.

\bibliographystyle{mnras}
\bibliography{WNJ_BB_aph}



\onecolumn

\appendix

\section{Comparison of the dynamics for the parameter runs}
\label{sect:para_dynamics}
Here, we show a more complete overview about certain variables resulting from our different parameter runs,
summarized in Table~\ref{tab:para_dynamics}.

\begin{figure*}
 \begin{center}
\includegraphics[height=2.5in,width=7.0in]{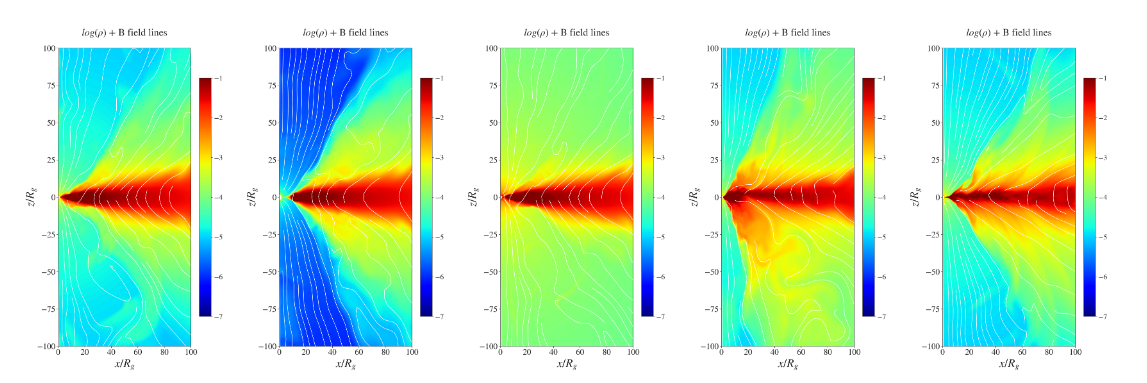}
\caption{Comparison of the density distribution for the parameter runs.
From left to right we show the mass density (in log10 scale), overlaid with poloidal magnetic field lines for
simulation runs 
{\em 21EF}, {\em 22EF}, {\em 23EF}, {\em 24EF}, {\em 26EF}, 
all at time $t=10000$, except {\em 26EF} shown at $t=4000$.
The density is given in code units (maximum disc density is unity).
Lengths are normalized in gravitational radii $R_{\rm g}$, while the time unit is $t_{\rm g}$.
}
\label{fig:para_dens}
\end{center}
\end{figure*}

Figure~\ref{fig:para_dens} shows the density structure of the disc-wind-jet system in comparison.
We see that the disc structure and also the disc wind look similar in all simulations, while the
Blandford-Znajek jet has a different mass loading.
The latter is set by the floor model of the MHD simulation. 
Simulation {\em 22EF} considers a non-rotating black hole.  
The low-density structure along the rotational axis is infalling material.
For simulation {\em 23EF} the floor model considers such a high density that a Blandford-Znajek jet, which requires
a sub-Alfv\'enic launching is not possible to drive.
The disc remains thin with a scale height somewhat little increase from its initial value.
In the figure, the {"}red{"} areas above the disc surface indicate a disc wind launched with relatively high density.

\begin{figure*}
 \begin{center}
\includegraphics[height=2.5in,width=7.0in]{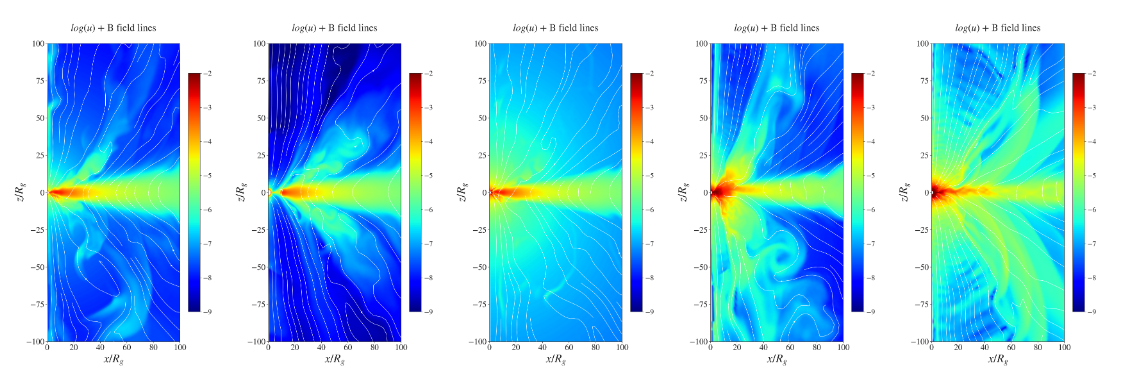}
\caption{Comparison of the internal energy distribution for the parameter runs.
From left to right we show the internal energy (in log10 scale) overlaid with velocity streamlines for
simulation runs 
{\em 21EF}, {\em 22EF}, {\em 23EF}, {\em 24EF}, {\em 26EF},
all at time $t=10000$, except {\em 26EF} shown at $t=4000$.
Lengths are normalized in gravitational radii $R_{\rm g}$, while the time unit is $t_{\rm g}$. 
Note that the simulation time for {\em 26EF} is shorter, as due to the strong magnetic field 
the (magneto)hydrodynamic wave is substantially faster while consuming the same CPU time.}

\label{fig:para_ener}
\end{center}
\end{figure*}

Figure~\ref{fig:para_ener} shows the internal energy distribution for the dynamical models in comparison.
Also here, the disc structure and the disc wind look comparable.
Dense gas, ejected from the disc surface into a disc wind also carries a relatively large internal energy.
This is in particular visible in simulation {\em 23EF}.
With the floor model, also a floor on the internal energy is set.
The infalling (floor) mass of simulation {\em 22EF} also has a low internal energy (thus temperature).

\begin{figure*}
 \begin{center}
\includegraphics[height=2.5in,width=7.0in]{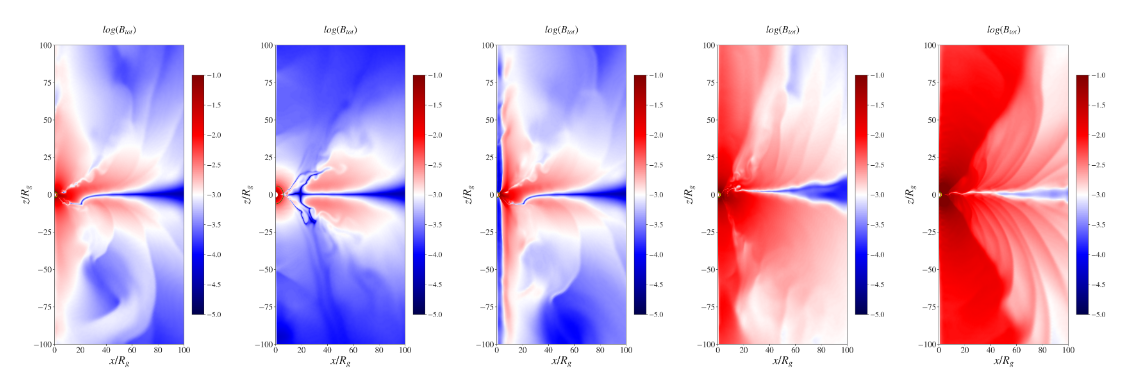}
\caption{Comparison of the magnetic energy distribution for the parameter runs.
From left to right we show the total magnetic energy (in log10 scale) for
simulation runs 
{\em 21EF}, {\em 22EF}, {\em 23EF}, {\em 24EF}, {\em 26EF},
all at time $t=10000$, except {\em 26EF} shown at $t=4000$.
Lengths are normalized in gravitational radii $R_{\rm g}$, while the time unit is $t_{\rm g}$. 
Note that the simulation time for {\em 26EF} is shorter, as due to the strong magnetic field 
the (magneto)hydrodynamic wave is substantially faster while consuming the same CPU time.}

\label{fig:para_btot}
\end{center}
\end{figure*}

Figure~\ref{fig:para_btot} shows the distribution of the total magnetic field in comparison for the different
dynamical models.
The magnetic field strength is primarily set by the choice of the initial plasma-$\beta$.
We clearly see the strong field of simulations {\em 24EF} and {\em 26EF}.
However, the magnetic flux carried by the accretion disc will also advected inwards towards the black hole
and can potentially increase the efficiency of the Blandford-Znajek jet.
Also the strength of the disc wind depends on the magnetic field strength (see Figure~\ref{fig:para_vel} in the main text).

\bsp	
\label{lastpage}
\end{document}